\documentclass[11pt]{article}

\usepackage{amsmath}%
\usepackage{amsfonts}%
\usepackage{amssymb}
\usepackage[]{graphicx}%
\usepackage{verbatim}%
\usepackage{multirow}%
\usepackage{array}%
\usepackage{multirow}
\usepackage{url}

\usepackage{authblk}
\usepackage[margin=25mm]{geometry}
\usepackage[font=small]{caption}

\usepackage[colorlinks=true,linkcolor=blue,citecolor=blue]{hyperref}

\providecommand{\keywords}[1]
{
  \small	
  \textbf{\textit{Keywords---}} #1
}

\newcommand{\eps}{\varepsilon}%
\newcommand{\Om}{\Omega}
\newcommand\J[4]{\frac{\partial(#1,#2)}{\partial(#3,#4)}}
\newcommand{\of}{OpenFOAM}

\newcommand{\alabel}[1]{\label{#1}}

\title{ {Numerical models for evolution of extreme wave groups} }
\author[$^1$]{Eugeny Buldakov}
\author[$^2$]{Pablo Higuera}
\author[$^3$]{Dimitris Stagonas}
\affil[{$^1$}]{UCL, Department of Civil Engineering, Gower Street, London, WC1E 6BT, UK}
\affil[{$^2$}]{NUS, Faculty of Engineering, Engineering Drive 2, 117578, Singapore}
\affil[{$^3$}]{UCL, Department of Mechanical Engineering, Gower Street, London, WC1E 6BT, UK}

\begin{document}

\maketitle

\begin{abstract}%
The paper considers the application of two numerical models to simulate the
	evolution of steep breaking waves. The first one is a Lagrangian wave
	model based on equations of motion of an inviscid fluid in Lagrangian
	coordinates. A method for treating spilling breaking is introduced and
	includes dissipative suppression of the breaker and correction of crest
	shape to improve the post breaking behaviour. The model is used to
	create a Lagrangian numerical wave tank, to reproduce experimental
	results of wave group evolution. The same set of experiments is
	modelled using a novel VoF numerical wave tank created using \of{}.
	Lagrangian numerical results are validated against experiments and VoF
	computations and good agreement is demonstrated. Differences are
	observed only for a small region around the breaking crest.
\end{abstract}%


%
\section{Introduction}%
%
Theoretical analysis and field evidence show that the largest waves belong to
	wave groups. From the analysis of Gaussian random processes
	\cite{Lindgren1970}, it follows that, in the linear approximation, the
	average profile of an extreme wave in a random Gaussian sea is a wave
	group, whose shape is proportional to the autocorrelation function of
	the random process. The theoretical results are confirmed by field
	measurements \cite{Phillips1993a,Christou2014}. Focussed
	wave groups were suggested as design waves \cite{Tromans1991} and are
	often used in computations of wave structure interaction
	\cite{Chen2014,Hu2016}.

The evolution of steep travelling wave groups includes complex physical
processes occurring over a wide range of time and space scales, e. g.
long-term nonlinear wave evolution and wave breaking. A numerical model capable
of adequately describing such processes should meet a wide range of sometimes
conflicting requirements. Advantages and disadvantages of different numerical
models make them suitable for modelling different aspects of the wave group
evolution. It is therefore natural to combine different models to model
different stages of wave evolution. This leads to the development of so-called
coupled or hybrid models. In hybrid models, a relatively simple and
computationally efficient model is used to simulate the long-term evolution of
a steep wave up until the initial development of the overturning crest. Results
of these calculations are then used as initial conditions for a model capable
to resolve the interaction of the overturning crest with the underlying water
surface. Similarly, for wave-structure interaction problems the far-field wave
evolution can be modelled by a fast model and the region near the structure by
a more sophisticated but computationally expensive model. This approach has now
received considerable attention and numerous hybrid models have been developed.
The most popular hybrid models usually couple a Boundary Element Method (BEM)
model for the far-field calculations and a Volume of Fluid (VoF) model for the
wave-structure interaction \cite{Kim2010,Sriram2014}.  Further examples include
the coupling of a Boussinesq model with an Smooth Particle Hydrodynamics (SPH)
model \cite{Narayanaswamy2010} and BEM with SPH \cite{Landrini2012}. A brief
literature review on the subject of models coupling can be found in
\cite{Sriram2014}.

An alternative method to describe water waves is using equations of fluid
motion in Lagrangian formulation. These equations are written in material
coordinates moving together with the fluid. Each material point of fluid
continuum is marked by a specific label, and the labels create a continuous set
of Lagrangian coordinates. In the Lagrangian description the free surface is
represented by a fixed boundary of the domain in Lagrangian coordinates and
equations of fluid motion in Lagrangian form to be solved in a fixed domain of
Lagrangian labels. Some numerical methods utilise certain elements of the
Lagrangian description. For example, in boundary element methods time
integration is often based on a mixed Eulerian-Lagrangian (MEL) formulation,
which traces fluid particles on the free surface \cite{Tsai1996,Grilli2001}.
The SPH approach can be considered as fully Lagrangian. In this method, the
fluid domain is represented by a set of material particles, which serve as
physical carriers of fluid properties. An integral operator with a compact
smoothing kernel is used to represent average properties of fluid at a certain
location. Each particle interacts with nearby particles from a domain specified
by the smoothing kernel \cite{Gomez-Gesteira2010,Violeau2016}. However, SPH
models do not refer to equations of fluid motion in Lagrangian coordinates and
therefore differ from methods in which the Lagrangian equations are directly
applied to solve water wave problems.

The first works using equations of fluid motion in Lagrangian formulation with
applications to water wave problems appeared in early 70's. The formulation
included kinematic equations of mass and vorticity conservation for internal
points of a domain occupied by an ideal fluid \cite{Brennen1970}. The dynamics
of the flow was determined by a free-surface dynamic condition. An alternative
approach used Navier-Stokes equations in material coordinates moving together
with fluid \cite{Hirt1970}. The next step was the development of an Arbitrary
Lagrangian-Eulerian (ALE) formulation \cite{Chan1975}. ALE uses a computational
mesh moving arbitrarily within a computational domain to optimise shapes of
computational elements and a problem is formulated in moving coordinates
connected to the mesh. For example, Eulerian (fixed mesh) or fully Lagrangian
(fluid moving mesh) formulations can be used in different areas.
Implementation of a finite element approach with irregular triangular meshes
for ALE formulation \cite{Braess2000} led to the development of a sophisticated
method capable of solving complicated problems with interfaces \cite{ALE:Book}.
However, ALE models are complicated both in terms of formulation and numerical
realisation and lack the main advantage expected of a Lagrangian method, namely
the simplicity of representing computation domains with moving boundaries. The
deformation of elementary fluid volumes remains comparatively simple for many
problems solved within the ideal fluid framework. Such problems include the
evolution of propagating wave groups and can be efficiently treated with
much simpler Lagrangian models, like the model described in \cite{Brennen1970}.

The current work focuses on the development of a new version of a fully
Lagrangian finite-difference model, which has been previously applied to
simulate tsunami waves in a wave flume \cite{Buldakov2013}, violent sloshing in
a moving tank \cite{Buldakov2014} and the propagation of wave groups on sheared
currents \cite{Buldakov2015,Chen2019}. The original Lagrangian model is
modified to include a dispersion correction term, which increases the order of
approximation of the dispersion relation and considerably reduces long-term
dispersion errors. To continue the calculations through the wave breaking, we
implement breaking suppression by local surface damping. It turns out, that
elimination of the crest overturning leads to errors in the wave shape after
the breaking. To reduce this error, we implement a procedure for crest shape
correction. The model demonstrates certain advantages over alternative methods
for modelling water waves, like Boundary Element Methods. In particular, BEM
methods use potential formulation and can not be used for rotational flows. The
Lagrangian model can be applied to flows with arbitrary distribution of
vorticity, which allows its application to problems with sheared currents
\cite{Buldakov2015,Chen2019}. This feature is also useful for the accurate
simulation of post breaking wave evolution since wave breaking results in
intense vortical motion. At the same time, the new Lagrangian model remains
relatively simple and can thus be optimised for high computational efficiency. 

The new version of the Lagrangian model can be considered as a good candidate
for the fast element of a hybrid model. For a hybrid model it is important
that model elements generate consistent results for surface elevation and
kinematics at the vicinity of high crests, when switching between the two
models takes place. To demonstrate the compatibility of models, we perform a
detailed comparison of surface elevation and kinematics between the Lagrangian
model and a VoF model, which is a typical candidate for a second component of a
hybrid model. We use an open source realisation of a VoF method implemented in
olaFlow, which is a numerical two-phase flow solver with a core based on \of{}.
The model includes a module for moving-boundary wave generation and absorption
\cite{Higuera2013,Higuera2015}. This extension allows simulation of piston and
flap type wave makers and, therefore, direct comparison between results
generated in different numerical models and in experiments using identical wave
generation methods. First, we validate both models against experimental
measurements of surface elevation of wave groups in a wave tank. Numerical wave
tanks (NWT) replicating the experimental set-up are created in each model and
the displacement of the experimental wavemaker is used as input for both NWTs.
This allows direct comparison between numerical and experimental
results and between the models. After validation, the Lagrangian model and
olaFlow are applied to simulate the evolution of a steep breaking wave group.
The surface elevation and flow kinematics computed with the Lagrangian model
and olaFlow are then compared. Particular attention is paid to the surface
elevation before, during and after the breaking onset and to the flow
kinematics under the high nearly breaking crest.


\section{Experimental data}%

Numerical results presented in this paper are validated against a set of
experimental data on the propagation of focussed wave groups obtained in the
wave flume of the Civil Engineering department at UCL. The flume is $45\,$cm
wide with a $12.5$\,m long working section extending between two piston-type
wavemakers. The wavemaker on the right end of the flume generates waves, while
the opposite paddle acts as a wave absorber. For all tests, the water depth
over a horizontal bed of the flume was set at $h=40$\,cm. The centre of the
flume is used as the origin of the coordinate system with the $x$-axis directed
towards the wave generator positioned at $x=6.25$\,m. The vertical $z$-axis
originates from the still water surface and is directed upwards. The wavemaker
is controlled by a force feedback system, which operates in the frequency
domain. Input of the control system is the linearised amplitude spectrum of the
generated wave. The control system uses discrete spectra and generates periodic
paddle motions. For our experiments we use an overall return period of
$64\,$sec, which is the time between repeating identical events produced by the
paddle. The range of frequencies used by the wave generator is from $1/64$ to
$2\,$Hz with $128$ equally-spaced discrete frequency components $f_n=n/64$\,Hz;
$n=1\dots128$. Overall, the system allows precise control of wave generation and
provides partial absorption of the incoming waves at the further end of the
flume. The surface elevation is monitored with a series of six resistance wave
probes and an ultrasonic sensor is used to record the paddle motion. A
schematic of the experimental layout can be seen in Figure~\ref{Layout}.

%
%
\begin{figure}[t!]
\centering
\resizebox{\textwidth}{!}{\includegraphics{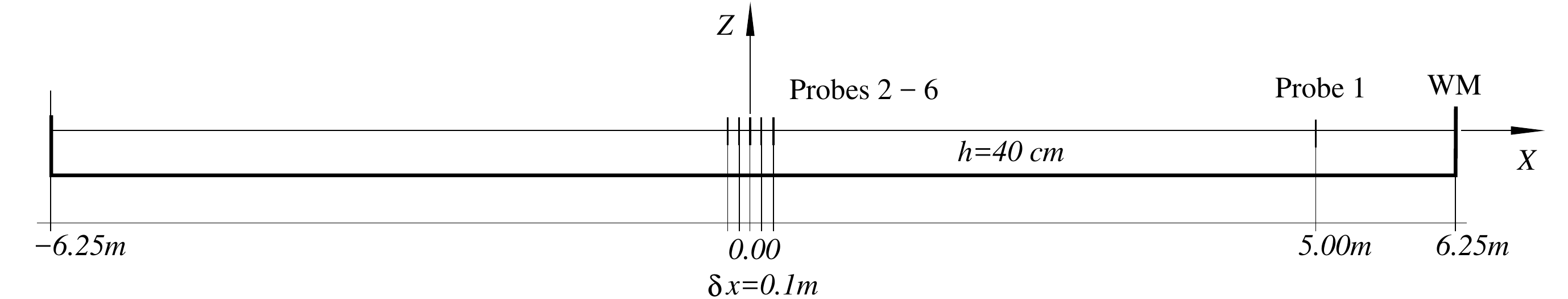}}
\caption{%
Wave flume layout and positions of wave probes.
}
\alabel{Layout}
\end{figure}
%
%

The wavemaker control does not account for dissipative and nonlinear effects,
and the spectrum of a generated wave group differs from the input spectrum of
the control system. The iterative procedure described in \cite{Buldakov2017} is
used to produce waves with the desired spectrum and focussed at the centre of
the flume. We apply the iterations to generate a Gaussian wave group with peak
frequency of $f_p=1$\,Hz focussed at the centre of the flume with the linear
focus amplitude of $2.5\,$cm. Then we use the resulting input spectrum to
generate higher amplitude waves by multiplying the input by factors of 2 and 3
without further corrections to phase or amplitude. We therefore obtain waves
with linear focus amplitudes $A=5\,$cm and $A=7.5\,$cm.  For all cases waves
with constant phase shifts $\Delta\phi=\pi/2$, $\pi$ and $3\pi/2$ were also
generated. For example, the phase shift $\Delta\phi=\pi$ corresponds to the
wavemaker input opposite to the original wave. Wave records with shifted phases
can be used for their spectral decomposition and linearisation
\cite{Buldakov2017}. The linearised spectra of generated waves at $x=0$ are
shown on Figure~\ref{EXP}. As expected, nonlinear defocussing and
transformation of the spectrum can be observed for higher amplitude waves.  The
resulting waves are of tree distinct qualitative types. The small amplitude
waves ($A=2.5\,$cm) have weakly nonlinear features. The waves with $A=5\,$cm
can be described as strongly nonlinear non-breaking waves. And the high
amplitude waves ($A=7.5\,$cm) exhibits intense breaking events as they travel
along the flume. 

Six return periods are generated in each experimental run. The first period
initiates from still water conditions and thus the surface elevation record
differs from the records for the following five return periods. The records for
these five periods demonstrate a high level of repeatability. It was therefore
assumed that the periodic wave system with the return period of $64$\,sec is
established in the flume after the first period. The data for the first return
period are neglected and the rest of the data are averaged between the
following five periods. This reduces contributions from all components with
periods other than the return period, including high frequency noise and
sloshing modes of the flume. Each of the six-period runs was repeated at least
three times, demonstrating a high level of repeatability.

%
%
\begin{figure}[t!]
\centering
\resizebox{0.9\textwidth}{!} 
{ 
\includegraphics{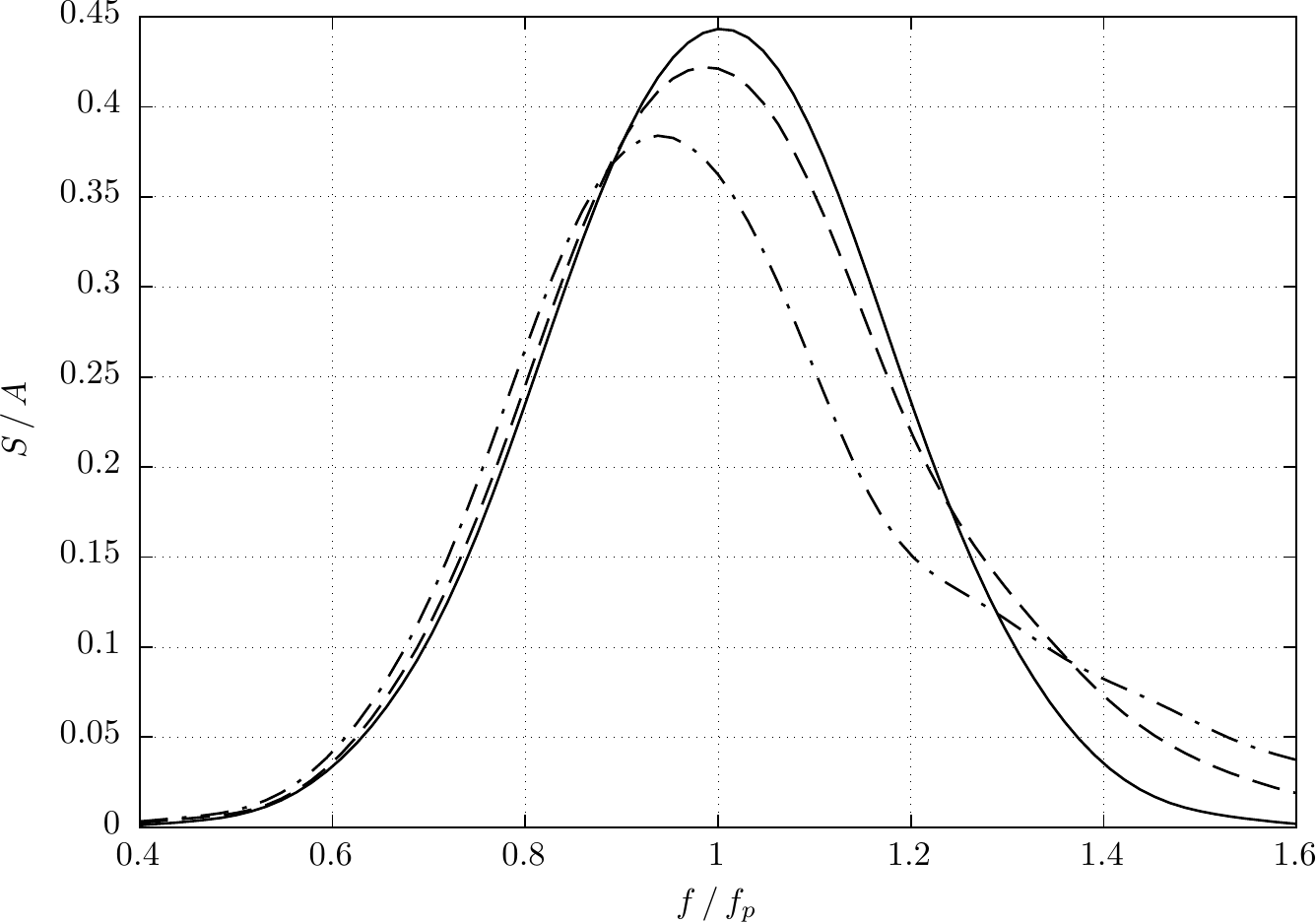}
\includegraphics{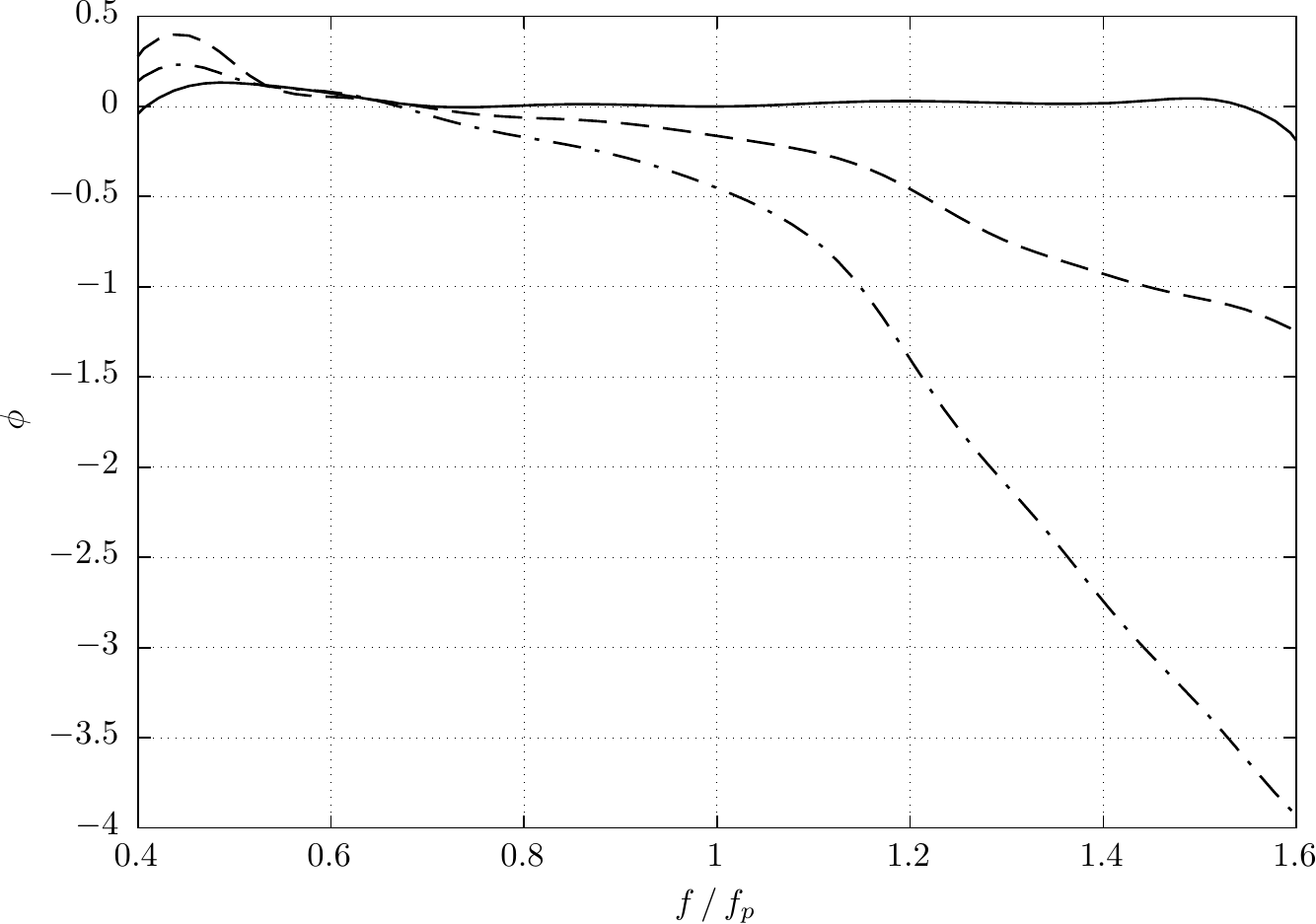}
}
\caption{
Linearised spectra of experimental wave groups at a focus point. Amplitude (left) and phase (right).
Solid-- $A=2.5$\,cm; dashed-- $A=5\,$cm; dash-dotted-- $A=7.5\,$cm.}
\alabel{EXP}
\end{figure}
%
%

\section{Lagrangian numerical model}%

	\subsection{Mathematical and numerical formulation}%

A general Lagrangian formulation for two-dimensional flow of inviscid fluid
with a free surface can be found in \cite{Buldakov2006}. We consider time
evolution of coordinates of fluid particles $x(a,c,t)$ and $z(a,c,t)$ as
functions of Lagrangian labels $(a,c)$. The formulation includes the Lagrangian
continuity equation and the Lagrangian form of vorticity conservation
\begin{equation}\label{eq:cont:vort}
\J{x}{z}{a}{c}=J(a,c)\,;\quad \J{x_t}{x}{a}{c}+\J{z_t}{z}{a}{c}=\Om(a,c)\,,
\end{equation}
and the dynamic free-surface condition
\begin{equation}\label{eq:dynamics}
x_{tt}x_a+z_{tt}z_a+g\,z_a \,\big|_{c=0}=0\,.
\end{equation}
Functions $J(a,c)$ and $\Om(a,c)$ are given functions of Lagrangian coordinates
and should be provided as a part of the problem formulation.

For the convenience of the numerical realisation, we modify the original
equations (\ref{eq:cont:vort}) and write them in the following form
\begin{equation}\alabel{eq:eqns}
  \Delta_{t}\left( \J{x}{z}{a}{c} \right) =0;\qquad \Delta_{t}\left(
	\J{x_t}{x}{a}{c}+\J{z_t}{z}{a}{c} \right)=0\,,
\end{equation}
where the operator $\Delta_{t}$ denotes the change between time steps. From the
point of view of the numerical realisation, equations (\ref{eq:eqns}) mean that
values in brackets at two time steps are equal to one another. This formulation
does not require explicit specification of functions $J(a,c)$ and $\Om(a,c)$,
which are specified implicitly by the initial conditions. $J(a,c)$ is defined
by initial positions of fluid particles associated with labels $(a,c)$ and
$\Om(a,c)$ is the vorticity distribution, which is defined by the velocity field
at the initial time. We also modify the dynamic free surface condition (\ref{eq:dynamics})
by adding various service and physical terms to the right-hand side
\begin{equation}\alabel{eq:surf}
	x_{tt}x_a+z_{tt}z_a+g\,z_a= {-RHS(a,t)}\,\Big|_{c=0}\,.
\end{equation}
We use the following set of additional terms
\begin{equation}\alabel{eq:rhs}
 RHS(a,t)= {k\,( x_{t}\,x_a+z_{t}z_a)}\,+\,
	{\sigma\,\frac{\partial}{\partial\,a}\,
	\frac{\partial\,\kappa}{\partial\,t} }\,+\,
	{\gamma\,\frac{\partial\kappa}{\partial a} }\,,
\end{equation}
where $\kappa$ is the surface curvature. The first term in (\ref{eq:rhs})
introduces damping of displacement of surface particles with the damping
coefficient $k(a)$. This term is used for absorbing reflections from a boundary
of a numerical wave tank opposite to a wavemaker. The second term represents
damping of surface curvature and $\sigma(a)$ is the corresponding damping
coefficient. As described later in this paper, it is used to simulate the
dissipative effects of wave breaking. The third term represents surface tension
with the coefficient~$\gamma$.

A specific problem within the general formulation is defined by boundary and
initial conditions. In this paper boundary conditions are used to simulate the
experimental wave tank presented on Figure~\ref{Layout}. We use a rectangular
Lagrangian domain with $c=0$ being the free surface and $c=-h$ being the
bottom. The known shape of the bottom provides the condition on the lower
boundary of the Lagrangian domain. For the case of a flat bed of depth $h$ we
have $z(x,-h)=-h$. On the right boundary of the Lagrangian domain $a=a_{\max}$
a moving vertical wall represents a piston wavemaker:
\begin{equation}\alabel{eq:wm}
x(a_{\max},c,t)=a_{\max}+X_\mathrm{wm}(t)\,, 
\end{equation}
where $X_\mathrm{wm}(t)$ is the prescribed wavemaker motion. A solid vertical
wall is used on the left boundary $x(a_{\min})=a_{\min}$. An absorption region
is implemented to reduce reflections from the wall. Still water initial
conditions are used to specify initial velocities and positions of fluid
particles. The set of equations (\ref{eq:eqns}) with the free surface boundary
condition (\ref{eq:surf},~\ref{eq:rhs}) and the conditions on the tank
boundaries is solved numerically using a finite-difference technique. 

Since equations (\ref{eq:eqns}) for internal points of a computational domain
include only first order spatial derivatives, a compact four-point Keller box
scheme \cite{Keller} can be used for finite-difference approximation of these
equations. For our selection of the Lagrangian domain the stencil
box can be chosen with sides parallel to the axes of the Lagrangian coordinates,
which significantly simplifies the final numerical scheme. The values
of the unknown functions $x$ and $z$ on the sides of the stencil box are
calculated as averages of values at adjacent points and then used to
approximate the derivatives across the box by first-order differences. The
scheme provides the second-order approximation for the box central point. Time
derivatives in the second equation in (\ref{eq:eqns}) are approximated by
second-order backward differences. Spatial derivatives in the free-surface
boundary condition are approximated by second-order central differences and
second-order backward differences are used to approximate time derivatives. As
demonstrated below, this leads to a numerical scheme with weak dissipation. The
overall numerical scheme is of the second order in both time and space.

A fully-implicit time marching is applied, and Newton method is used on each
time step to solve the nonlinear algebraic difference equations. The scheme
uses only 4~mesh points in the corners of the box for internal points of the
fluid domain. Therefore, the resulting Jacobi matrix used by nonlinear Newton
iterations has a sparse 4-diagonal structure and can be effectively inverted
using algorithms, which are faster and less demanding for computational memory
than general algorithms of matrix inversion. The current version of the solver
uses a standard routine for inversion of general sparse matrices \cite{NAG}. To
reduce calculation time, the inversion of a Jacobi matrix is performed at a
first step of Newton iterations and if iterations start to diverge. Otherwise,
a previously calculated inverse Jacobi matrix is used. Usually only one matrix
inversion per time step is required. To start time marching, positions of fluid
particles at three initial time steps should be provided, which specifies
initial conditions for both particle positions and velocities. An adaptive mesh
is used in the horizontal direction with an algorithm based on the shape of the
free surface in Lagrangian coordinates $z(a,0,t)$ to refine the mesh in regions
of high surface gradients and curvatures. Constant mesh refinement near the
free surface is used in the vertical direction.

Since the finite-difference approximation presented above uses quadrangular
mesh cells, it may be subjected to the so called alternating errors caused by
non-physical deformations of cells \cite{Hirt1970,Chan1975}. This deformation
preserves the overall volume (area) of a cell, as prescribed by finite
difference approximation of the Lagrangian continuity equation. However,
volumes of triangles build on cell vertexes can change. The area of one of the
triangles increases while the area of the second triangle decreases by the same
value. 
%
%
This leads to alternating 
%
%
distortions of neighbouring cells, resulting in instability of short-wave
disturbances with a wavelength equal to two cell spacing. It is usual to
implement artificial smoothing to suppress this instability \cite{Chan1975}.
For the presented model, efficient suppression of the instability caused by
alternating errors is provided by an adaptive mesh. After several time steps
the solution is transferred to a new mesh using quadratic interpolation. The
effect of this procedure is similar to regridding used in
\cite{Dommermuth1987}. It reduces short-wave disturbances and does not produce
unwanted damping. Alternating errors remain the main reason for calculations
breakdown. However, this happens for large deformations of the domain
associated, for example, with wave breaking. Otherwise, the numerical scheme
is stable with respect to this type of instability

	\subsection{Numerical dispersion and dispersion correction}%

For travelling waves dispersion errors can be the main source of numerical
errors. Therefore, special attention must be paid to approximation of the
dispersion relation. To derive the numerical dispersion relation we use the
continuous representation of the finite difference scheme. In this method the
finite difference operators are expanded to Taylor series with respect to small
discretisation steps. For example, a 3-point backward difference approximation
of the second time derivative can be expressed as
\begin{equation}\alabel{eq:findiff}
	\begin{split}
	\frac{-f(t-3\,\tau)+4\,f(t-2\,\tau)-5\,f(t-\tau)+2\,f(t)}{\tau^2}= \qquad\qquad\qquad \\
		=f''(t)-\frac{11}{12}\,\tau^2\,f''''(t)+O(\tau^3)\,.
	\end{split}
\end{equation}
As can be seen, the approximation is of the second order with the leading term
of the error proportional to the fourth derivative of a function, which gives
the main contribution to the error. Similar
continuous representations can be derived for all other difference operators. 

We consider the spatial and temporal discretisation according to the numerical
scheme described above with the time
step $\tau$ and spatial steps $(\delta a; \delta c)$. Under an assumption of
small perturbations of the original particle positions, we represent unknown
functions in the form
\[
x=a+\eps\,\xi(a,c,t);\quad z=c+\eps\,\zeta(a,c,t)
\]
and keep only linear terms of expansions with respect to the small displacement
amplitude $\eps\to 0$. The unknown functions $\xi$ and $\zeta$ represent
displacements of fluid particles from the original positions. The solution for
a regular travelling wave in deep water can be written as
\[
\xi=i\,e^{i\,k\,a}\,e^{\varkappa\,c}\,e^{i\,\omega\,t}\,;\quad
\zeta=e^{i\,k\,a}\,e^{\varkappa\,c}\,e^{i\,\omega\,t}\,.
\]
Because of the discretisation error, the constant for the exponential decay of
the displacement with the depth $\varkappa$ is not equal to the wavenumber $k$.
We are looking for the solutions for $\omega$ and $\varkappa$ as expansions
with respect to small discretisation parameters. The expansion for $\varkappa$
should satisfy the discrete version of (\ref{eq:cont:vort}).
The expansion for $\omega$ defines the numerical dispersion relation and is
used to satisfy the free-surface condition (\ref{eq:dynamics}). The
corresponding expansions are found to be
\begin{equation}\alabel{eq:kappa}
\varkappa/k=1-\frac{1}{24}\,(\, \delta\hat{a}^2+\delta\hat{c}^2 \,) +
	O(\delta\hat{a}^4;\delta\hat{c}^4)
\end{equation}
and
\begin{equation}\alabel{eq:numdis-0}
\omega/\sqrt{gk}=\,\pm1 \mp\frac{1}{24}\,(11\,\hat{\tau}^2+2\,\delta \hat{a}^2)
+\frac{1}{2}\,i\,\hat{\tau}^3+O(\hat{\tau}^4;\delta\hat{a}^4)\,,
\end{equation}
where the nondimensional discretisation steps $\hat\tau=\sqrt{gk}\,\tau$,
$\delta\hat a=k\,\delta a$ and $\delta\hat{c}=k\,\delta c$ are used. 
As can be seen, the numerical dispersion relation (\ref{eq:numdis-0})
incorporates a numerical error to dispersion at the second order
and has a weak dissipation at the third order. Equation
(\ref{eq:kappa}) shows that there is also a second-order error in the
displacement of fluid particles. It is interesting to note that the dispersive
error is affected only by the horizontal discretisation step, while the
vertical discretisation affects the kinematics of fluid particles under the
surface. Therefore, if we are interested in the evolution of the waveform, we
can use relatively few vertical mesh points.

Validation tests show that the achieved convergence rate for the
second-order dispersion approximation is not sufficient. To increase the
approximation order for the dispersion relation we introduce dispersion
correction terms to the free surface boundary condition. These terms should
satisfy the following conditions: (i) to have the order of $O(\tau^2;\delta
a^2)$; (ii) to be linear; (iii) not include high derivatives; (iv) to use the
same stencil as the original scheme and (v) to reduce the order of the
dispersion error. It has been found that the free surface boundary condition
(\ref{eq:dynamics}) with the dispersion correction term satisfying these
conditions can be written as follows
\begin{equation}\alabel{eq:discor}
x_{tt}x_a+z_{tt}z_a+g\,z_a+ \left( \frac{1}{6}\delta
	a^2\,x_{aa,tt}-\frac{11}{12} g\,\tau^2 z_{a,tt} \right) =0\,
	\Big|_{c=0} \,,
\end{equation}
where the dispersion correction term is given in brackets. The term $x_{aa}$
with the second-order spatial derivative leads to high-wavenumber nonlinear
instability for large wave amplitudes. To suppress this instability, we apply
5-point quadratic smoothing to function $x$ at the surface before applying the
finite-difference operator. The smoothing is applied only to this term in
(\ref{eq:discor}). It is applied at the future time layer to ensure that the
scheme remains fully implicit. It can be shown that the numerical dispersion
relation becomes
\begin{equation}\alabel{eq:numdis}
\omega/\sqrt{gk}=\,\pm1 +\frac{1}{2}\,i\,\hat{\tau}^3
	\pm\,(\frac{361}{480}\hat{\tau}^4-\frac{1}{360}\delta \hat{a}^4)
	-\frac{13}{12}\,i\,\hat{\tau}^5 +O(\hat{\tau}^6;\delta\hat{a}^6)\,.
\end{equation}
We now have weak numerical dissipation at 3-rd order and the dispersion error
at 4-th order. A term with weak negative dissipation at 5-th order should also
be noted. This term can potentially lead to solution instability for large time
steps.

	\subsection{Treatment of breaking}%

\begin{table*}[t!]%
\centering
	\begin{tabular}{|c|c|c|c|c||c|c|c|c|c|}
\hline
$N_x$ & \multicolumn{3}{c|}{$N_z$} & $\delta t$,\,sec &
$N_x$ & \multicolumn{3}{c|}{$N_z$} & $\delta t$,\,sec \\
\hline 
		101 & 11 & 16 & 21 & 0.010	&
		251 & 11 & 16 & 21 & 0.004	\\
\hline 
		126 & 11 & 16 & 21 & 0.008	&
		401 & 11 & 16 & & 0.0025	\\
\hline 
		201 & 11 & 16 & 21 & 0.005	&
		501 & 11 & & & 0.002	\\
\hline 
\end{tabular}
\caption{Numerical cases for convergence and validation tests of the Lagrangian model.}
\alabel{tbl}
\end{table*}%
%

A disadvantage of the Lagrangian model, BEM models and other simple models of
wave propagation is their inability to model spilling breaking. Temporal and
spatial scales of evolution of a breaking crest are small compare to a typical
wave period, wave length and wave amplitude. This leads to a singularity in a
numerical solution, which causes breakdown of computations. The Lagrangian
model can simulate overturning waves and therefore with sufficient spatial and
temporal resolution it can resolve micro-plungers originating at wave crests
during the initial stages of spilling breaking. However, the model can not
continue calculations after the self-contact of the free surface has occurred
and the solution becomes nonphysical. This makes it impossible to apply the
model to steep progressive waves and extreme sea conditions. Removing a
singularity in the vicinity of the breaking crest can help continuing
calculations with only a minor effect on the overall wave behaviour. This can
be achieved by implementing artificial local dissipation in the vicinity of a
wave crest prior to breaking. With this approach all small-scale local features
would disappear from the solution, but the overall behaviour of the wave would
be represented with good accuracy. Practical implementation of the method
employs a breaking criterion to initiate dissipation right before breaking
occurs. The dissipation is usually enforced by including damping terms in the
free-surface boundary conditions
\cite{Subramani1998,Guignard2001}. Recently, a breaking model
based on an advanced breaking criterion and an eddy viscosity dissipation model
was implemented in a spectral model of wave evolution
\cite{Tian2012,Seiffert2018}. The method demonstrates a good comparison with
the experiments and allows to apply spectral models to simulate evolution of
severe sea states with breaking.

In this paper we use a method of treatment for spilling breaking, which uses the
same basic concept but differs in its details of realisation. The method
includes dissipative suppression of the breaker and correction of crest shape
to provide accurate post-breaking behaviour of the wave. There are several
conditions such a method should satisfy: (i)~to act locally in the close
vicinity of a developing singularity without affecting the rest of the flow;
(ii)~to simulate energy dissipation caused by breaking; (iii)~to be
mesh-independent, that is the change of the effect with changing mesh
resolution should be within the accuracy of the overall numerical approximation
and (iv)~to be naturally included into a problem formulation representing an
actual or artificial physical phenomenon. The development of a spilling breaker
is associated with a rapid growth of surface curvature. Therefore, the local
dissipation effect satisfying these conditions can be created by adding a term
$-\sigma\,\partial/\partial a\,(\,\partial \kappa/\partial t \,)$ 
to the right hand side of the free surface dynamic condition
(\ref{eq:surf},~\ref{eq:rhs}). This term with a small coefficient $\sigma$
introduces artificial dissipation due to the change of surface curvature
$\kappa$, which acts locally at the region of fast curvature changes and
suppresses breaker development without affecting the rest of the wave. To
minimise the undesirable effect of dissipation, the action of the damping term
is limited both in time and in space. Breaking dissipation is triggered when
the maximal acceleration of fluid particles at the crest exceeds a specified
threshold $a_\mathrm{on}$ and it goes off when the maximum acceleration falls
below a second lower value $a_\mathrm{off}$. The appropriate values for
activation and deactivation thresholds are $a_\mathrm{on}=g$ and
$a_\mathrm{off}=0.5\,g$. Spatially, the action of the breaking model is limited
by the half-wavelength between the ascending and descending crossing points
around a breaking wave crest.

%
%
\begin{figure}[t!]
\centering
	\includegraphics[width=0.45\textwidth]{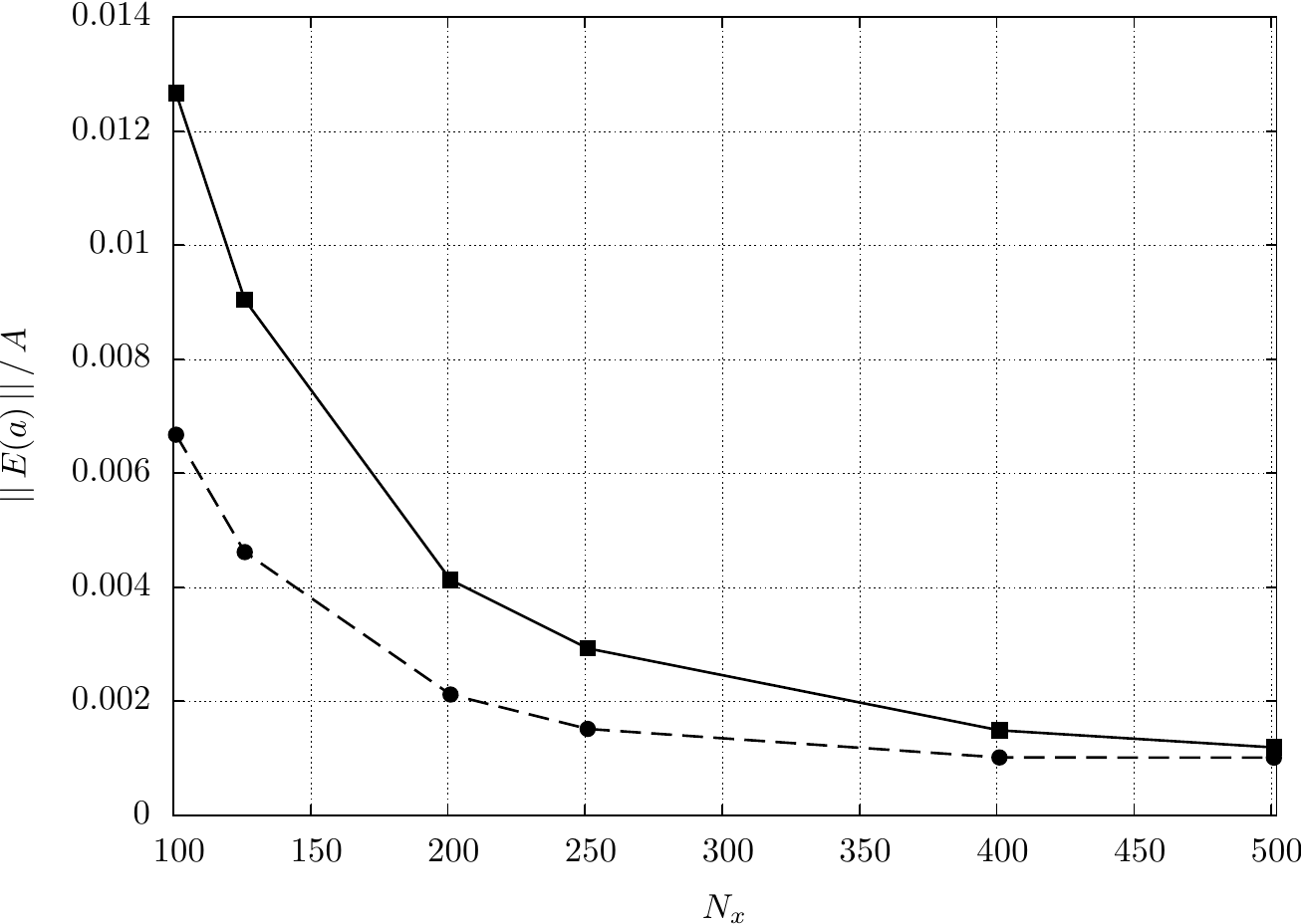}~
	\includegraphics[width=0.45\textwidth]{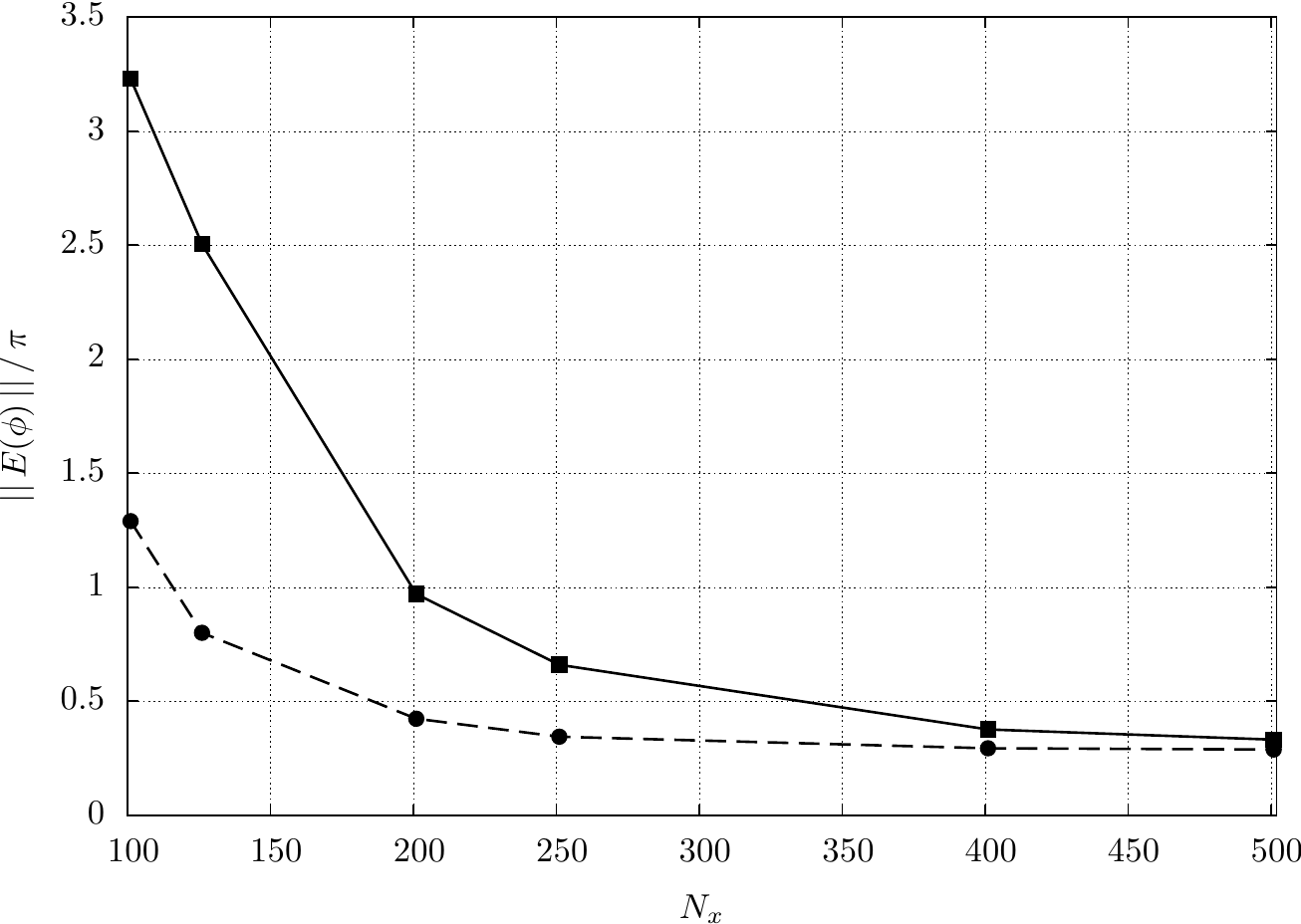}~~~\\[0.3em]
	\includegraphics[width=0.45\textwidth]{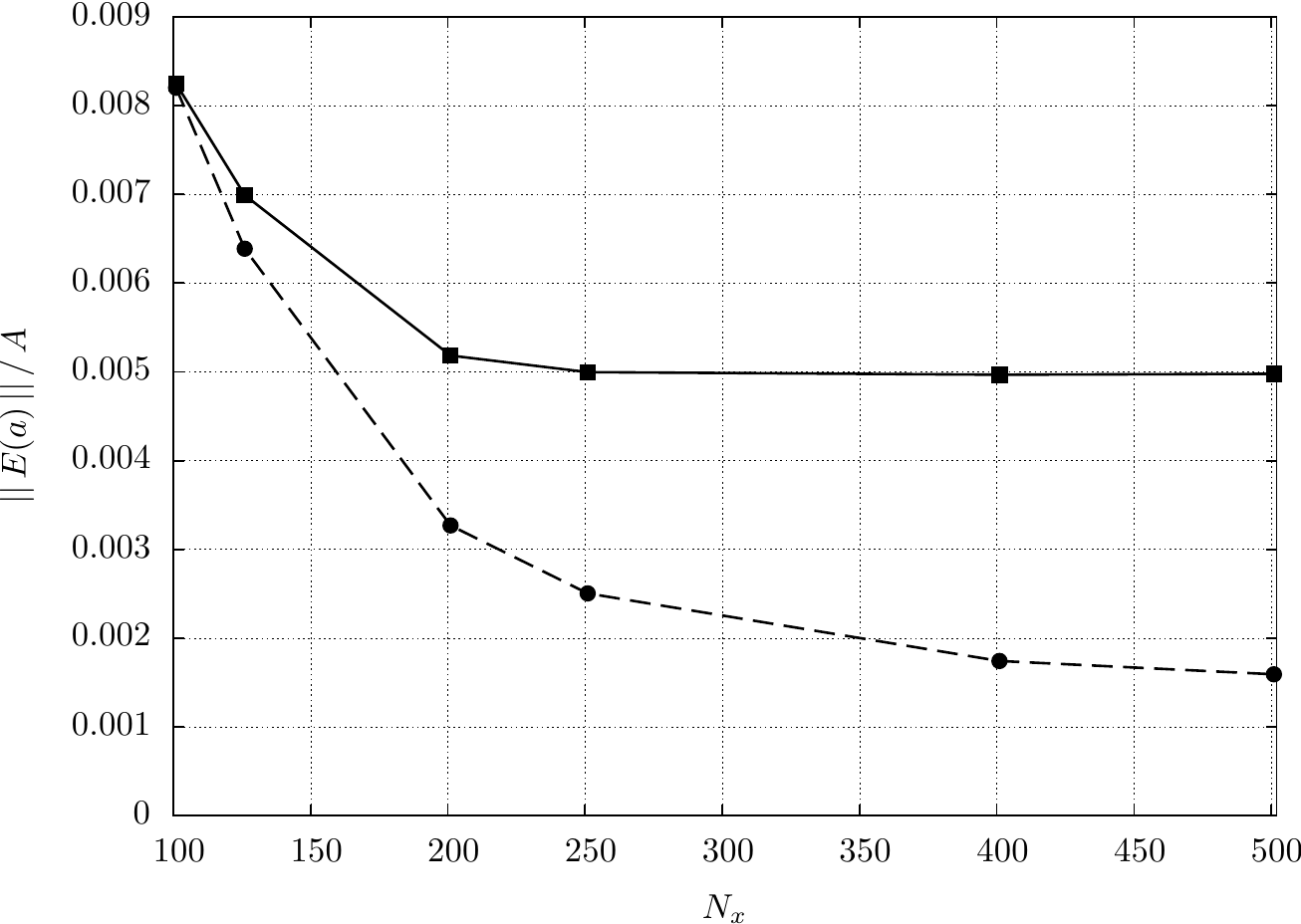}~
	\includegraphics[width=0.45\textwidth]{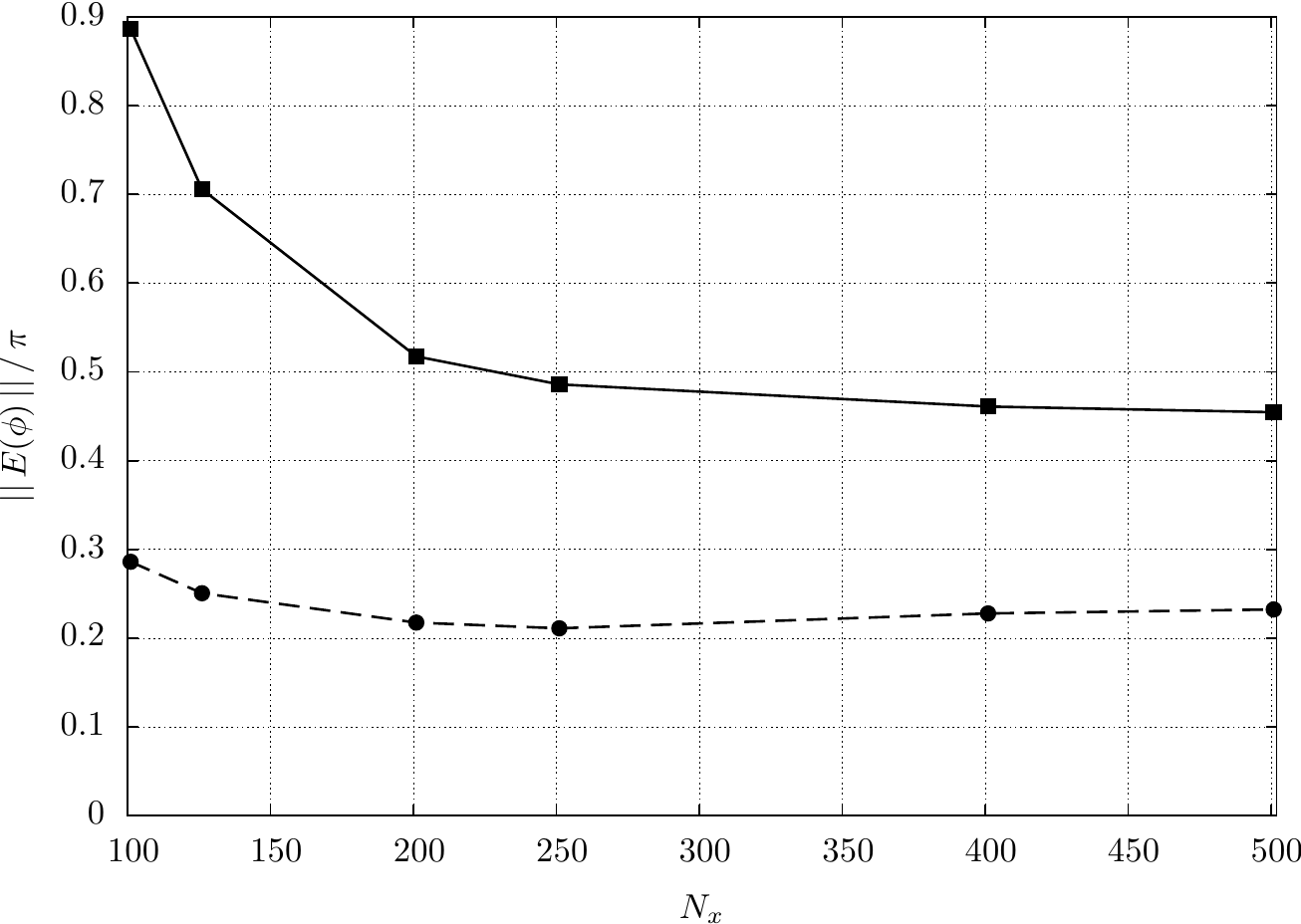}~~~
\caption{%
$L_2$-norm of difference between amplitude (left) and phase (right) spectra of
	a wave group at $x=0$ measured in experiment and calculated by the
	Lagrangian solver. Top -- $A=5$\,cm; without (solid) and with (dashed)
	dispersion correction. Bottom -- $A=7.5$\,cm; breaking control without
	(solid) and with (dashed) crest correction. $N_z=11$.
}
\alabel{fig:err}
\end{figure}
%
%

Activation of the damping term makes it possible to continue the calculation
beyond the breaking event. However, the resulting shape of the wave crest is
different from the actual one. Since the dissipation suppresses breaking,
overturning of the crest does not occur. For a sufficiently intense breaking,
it can significantly affect the shape of the entire wave around the crest. To
account for this difference, we apply surface tension around the crest, which
produces an effect similar to that of the peak overturning and reduces the
error in the shape of the post-breaking wave. This is achieved by adding the
surface tension term to the right side of the free surface boundary condition
(\ref{eq:surf},~\ref{eq:rhs}) with an appropriate coefficient
$\gamma=\gamma_\mathrm{br}$. It should be emphasised that this is not the
physical surface tension and that the desired effect is only possible if the
coefficient is much larger than the physical one. Natural surface tension can
also be implemented by selecting an appropriate value of $\gamma$. The large
$\gamma_\mathrm{br}$ required by the breaking model is used in the regions and
during the periods when the model operates. The intensity of the dissipation
($\sigma$), the acceleration thresholds to activate and deactivate the
dissipation ($a_\mathrm{on}$ and $a_\mathrm{off}$) and the strength of the
surface tension for the correction of the shape of the crest
($\gamma_\mathrm{br}$) constitute the four parameters of the breaking model. 
%

	\subsection{Model validation and convergence tests}
	\label{sec:LagrangianValidation}

%
%
\begin{figure}[t!]
\centering
	\includegraphics[width=0.9\textwidth]{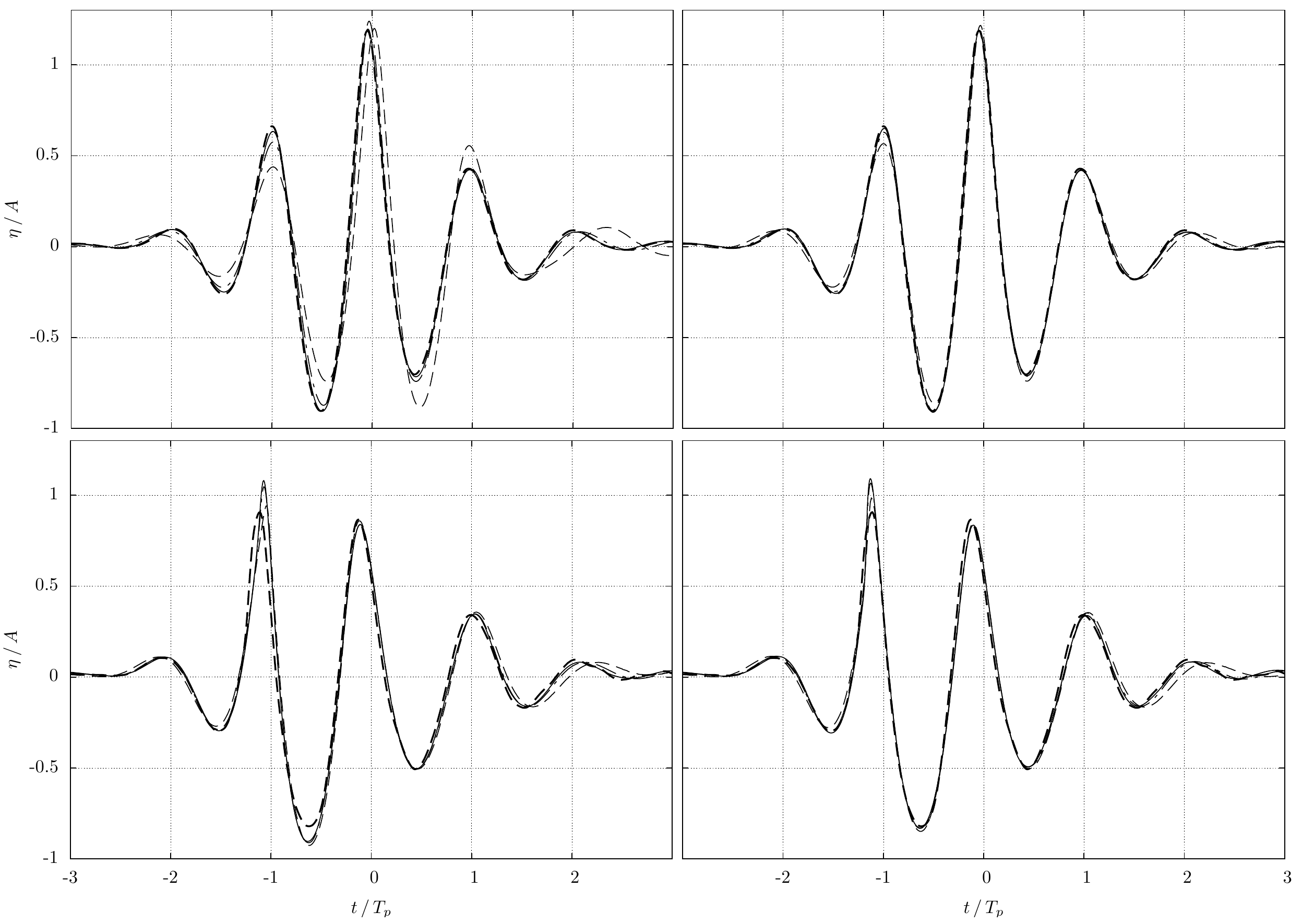}
\caption{%
Convergence of time history of surface elevation for a wave group at $x=0$. Experiment (thick
	dashed) and Lagrangian solver: $N_x=101$ (dashed); $N_x=201$
	(dash-dotted); $N_x=401$ (solid); $N_z=11$. Top -- $A=5$\,cm without
	(left) and with (right) dispersion correction. Bottom -- $A=7.5$\,cm,
	breaking control without (left) and with (right) crest correction.
} 
\alabel{fig:conv}
\end{figure}
%
%

We reconstruct the experimental setup using a numerical wave tank based on the
previously described Lagrangian numerical model. The dimensions of the NWT are
the same as the dimensions of the experimental tank. The wave is generated by
implementing the boundary condition (\ref{eq:wm}) on the right boundary of the
computational domain with the paddle displacement recorded in the experiment.
To account for gaps between the paddle and walls and bottom of the experimental
flume we reduce the actual paddle displacement by $18.5\%$. The surface
displacement damping term in (\ref{eq:rhs}) is activated near the tank wall
opposite the wavemaker to absorb the reflections. Calculations are performed
for all experimental cases, including three amplitudes and 4 phase shifts, and
repeated with different time steps and horizontal and vertical numbers of mesh
points. The time step has been reduced with an increasing number of horizontal
mesh points to maintain the dispersion errors due to temporal and spatial
discretisation given by (\ref{eq:numdis}) approximately equal to each other.
This provides uniform convergence by both parameters. The summary of parameters
for numerical cases is given in Table~\ref{tbl}.

The convergence is tested using a $L_2$ norm of the difference between the
experimental and calculated spectral components of the surface elevation at the
linear focus point $x=0$. The norms for spectral amplitudes and phases are
calculated as
\begin{equation}\alabel{eq:L2}
	||E(a)||=\sqrt{\sum_n(a_n-a_{n_\mathrm{EXP}})^2}\,;\quad
	||E(\phi)||=\sqrt{\sum_n(\phi_n-\phi_{n_\mathrm{EXP}})^2}\,,
\end{equation}
where the sum is taken over discrete spectral components in the range
$0.5\,\mathrm{Hz}<f<1.5\,\mathrm{Hz}$ from the set generated by the wavemaker ($n=32\dots96$),
where amplitude components are large enough not to be affected by experimental
errors. It should be noted that full convergence of the calculated results to
the experimental measurements can not be expected. The numerical model is based
on a set of assumptions that are satisfied with limited precision. In addition,
the measurements are also subject to accuracy limitations and experimental
errors. Therefore, the difference between the experimental and numerical
results converges to a certain small value and does not change with an
additional increase in the resolution of the numerical model.

%
%
\begin{figure}[t!]
\centering
\includegraphics[width=0.9\textwidth]{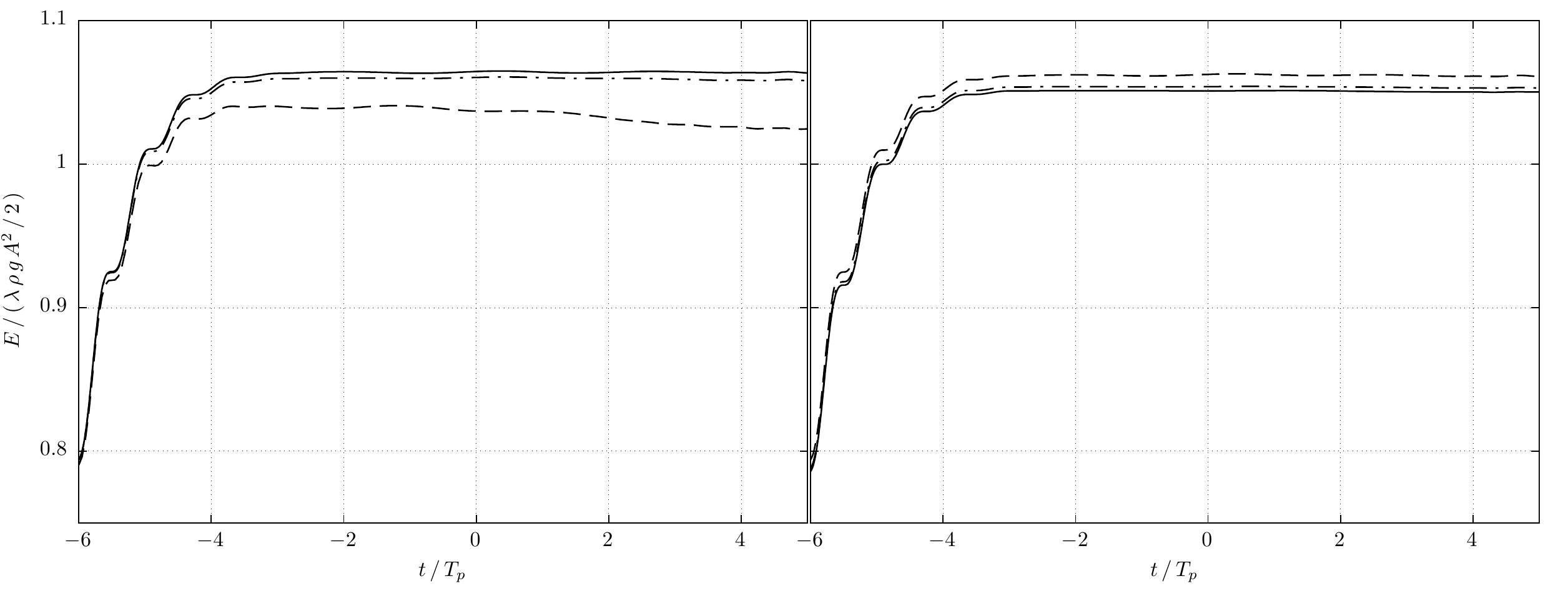}~~
\caption{%
Convergence of the normalised total energy for the
Lagrangian solver with dispersion correction. $A=5$\,cm.
	Left -- increasing horizontal mesh resolution:
	$N_x=101$ (dashed); $N_x=201$ (dash-dotted); $N_x=401$ (solid); $N_z=11$.
	Right -- increasing vertical mesh resolution:
	$N_z=11$ (dashed); $N_z=16$ (dash-dotted); $N_z=21$ (solid); $N_x=251$.
}
\alabel{fig:engconv}
\end{figure}
%
%

Selected results for the convergence tests are presented in
Figures~\ref{fig:err}-\ref{fig:engconv}. As can be seen in the top row of
Figure~\ref{fig:err}, the results with and without dispersion correction
converge to the same solution. However, the introduction of a dispersion
correction considerably increases the speed of convergence for both amplitudes
and phases. The experimental results can be reproduced with sufficient accuracy
for a relatively small number of horizontal mesh points and a relatively large
time step. For a breaking wave (Figure~\ref{fig:err}, bottom row), the shape
correction of the crest introduces a new physical process. For this reason, the
numerical results with and without crest correction converge to different
solutions, and the numerical result with the correction shows a much better
comparison with the experiment. A general impression of convergence and
accuracy of the different versions of the numerical model can be obtained from
the graphs of the time history of surface elevation presented in
Figure~\ref{fig:conv}. 

Figure~\ref{fig:engconv} shows the behaviour of the total wave energy in the
wave tank calculated for different temporal and horizontal resolutions and
different resolutions of the vertical mesh. Reflection absorption is disabled
for the energy tests. The energy is normalised by the energy of one wave length
$\lambda$ of a linear regular wave with a frequency and amplitude equal
to the peak spectral frequency and the linear amplitude of the focussed wave.
The kinetic energy is calculated by numerical integration over the entire
Lagrangian fluid domain using a bi-linear interpolation of the velocities of
the fluid particles inside mesh cells. This provides the second order
approximation of the integral with respect to the mesh resolution. According
to (\ref{eq:kappa}), the error of the velocity profiles within the fluid
domain, and thus the kinetic energy, is determined by the horizontal and
vertical resolution of the mesh. The potential energy is calculated as the
integral of the potential energy density on a free surface and it is not directly
affected by the discretisation of the vertical mesh. As can be seen in the
figure, the total energy in the tank increases up to $t/T_p\approx -4$
due to the energy generated by the wave paddle. For low horizontal mesh
resolution, the dissipative term in the numerical dispersion relation leads to
the reduction of the energy of the propagating wave. For higher resolutions,
energy conservation is satisfied with high accuracy. The right graph of
Figure~\ref{fig:engconv} shows the rapid convergence of energy with increased
vertical resolution. This includes both the convergence of the numerical
integral used to calculate kinetic energy and the convergence of the numerical
solution for wave kinematics, as indicated by expansion (\ref{eq:kappa}).
Overall, the convergence tests show that the numerical results converge towards
a solution, which approximates the experiment with a good accuracy. The
implementation of the dispersion correction term greatly increases the
convergence rate and provides accurate results with a smaller number of mesh
points and a larger time step.

\section{VoF model}%

In this paper we also use olaFlow \cite{olaflow}, an open source Navier--Stokes
model developed as a continuation of the work in \cite{higuerat}. The model is
based on \of{} \cite{weller98}. It includes enhancements to generate and absorb
waves with moving boundaries \cite{Higuera2015}, which allows to replicate
piston and flap type wavemakers.

	\subsection{Mathematical formulation}

This computational fluid dynamics (CFD) model solves the three-di\-men\-sion\-al
Reynolds-Averaged Navier-Stokes (RANS) equations, for two incompressible phases
(water and air). Both fluids are considered a continuum mixture in which free
surface is tracked with the Volume of Fluid (VoF) technique \cite{hirt81}.
The model solves the conservation of mass and momentum equations:
\begin{equation} \label{eq:contOF}
	\nabla \cdot ( \rho \, \bf{U}) = 0
\end{equation}
\begin{eqnarray} \label{eq:momConsOF}
		\frac{\partial \rho \bf{U}}{\partial t} + \nabla \cdot (\rho
		\bf{UU}) = 
		- \nabla \it{p}^{*} - \bf{g} \cdot \bf{X} \, \nabla \rho 
		+ \nabla \cdot (\mu \nabla \bf{U} - \rho \overline{\bf{U}'
		\bf{U}'}) 
		+ \sigma \kappa \bf{\nabla \alpha}
\end{eqnarray}
%
%
in which $\rho$ is the density of the fluid, $t$ is time, $\nabla$ is the
vector differential operator $(\partial/\partial x, \partial/\partial y,
\partial/\partial z)$ and $\bf{U}$ is the Reynolds averaged velocity vector,
$p^{*}=p-\rho \, \bf{g} \cdot \bf{X}$ is the dynamic pressure, 
$\bf{g}$ is the acceleration due to gravity and $\bf{X}$ is the
position vector.

The molecular dynamic viscosity of the fluid ($\mu$) and the Reynolds stresses
($\rho \overline{\bf{U}' \bf{U}'}$) constitute the viscous term.  In RANS, the
Reynolds stresses can be modelled by different turbulence closure models, which
introduce a dynamic turbulent viscosity ($\mu_t$).  Generally, the viscous term
is written in terms of the effective dynamic viscosity $\mu_{\mbox{\tiny eff}}
= \mu + \mu_t$.  The last term in equation (\ref{eq:momConsOF}) introduces the
surface tension force \cite{brackbill92}.  In this term, $\sigma$ is the surface
tension coefficient, $\kappa$ is the curvature of the free surface, calculated
as $\nabla \cdot {\bf{\nabla \alpha}}\,/\,{|\bf{\nabla \alpha}|}$, and $\alpha$
is the Volume of Fluid (VoF) indicator function, introduced next.

In VoF, a continuous indicator function $\alpha$ represents the amount of water
per unit volume in a cell.  Thus, 1 is a pure water cell, 0 is a pure air cell,
and $0 < \alpha <1$ belongs to the interface.  The movement of the fluids is
tracked by a conservative advection equation, which needs to be bounded between
0 and 1, to be conservative and to maintain a sharp interface.  The VoF
equation in \of{} is as follows:
\begin{eqnarray} \label{eq:vofOF}
		\frac{\partial \alpha}{\partial t} + \nabla \cdot (\alpha
		\bf{U})
		+ \nabla \cdot [\alpha (1-\alpha) \bf{U}_c] = 0
\end{eqnarray}
The last term in (\ref{eq:vofOF}) is an artificial compression (anti-diffusion) 
term to prevent the spread of the interface
\cite{berberovic09}, which occurs because the numerical solutions of advection
equations often suffer from diffusion.  Within this term, $\bf{U}_c$ is a
compression velocity in the normal direction to the free surface, acting only
at the interface. The reader is referred to \cite{higuerat} for a complete
description.  Since the fluid mixture can be defined with $\alpha$, any fluid
properties like density ($\rho$) or viscosity ($\nu$,$\mu$) are calculated as a
weighted average, e.g.:
\begin{eqnarray} \label{eq:rhoVoF}
		\rho = \alpha \, \rho_w + (1-\alpha) \, \rho_a
\end{eqnarray}
where subindices $w$ and $a$ denote water and air, respectively.

%
%
\begin{figure}[t!] \centering
\resizebox{0.9\textwidth}{!}{\includegraphics{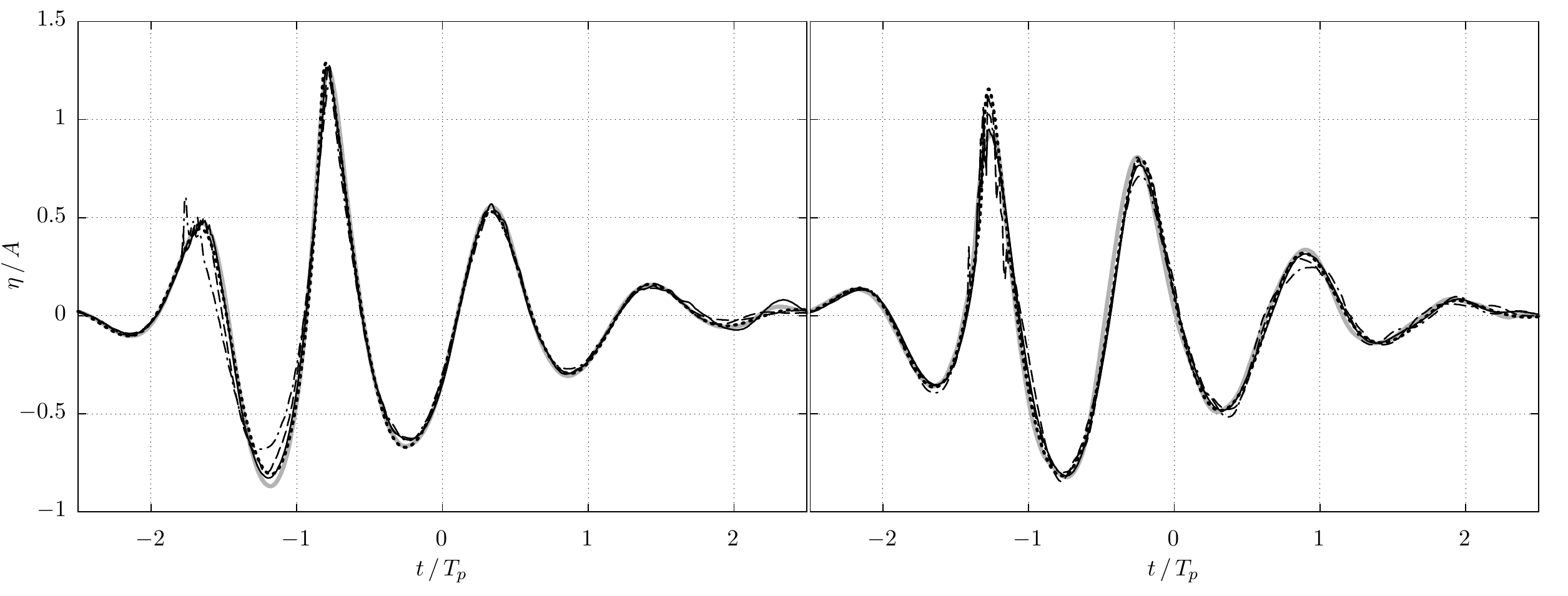}}
\resizebox{0.9\textwidth}{!}{\includegraphics{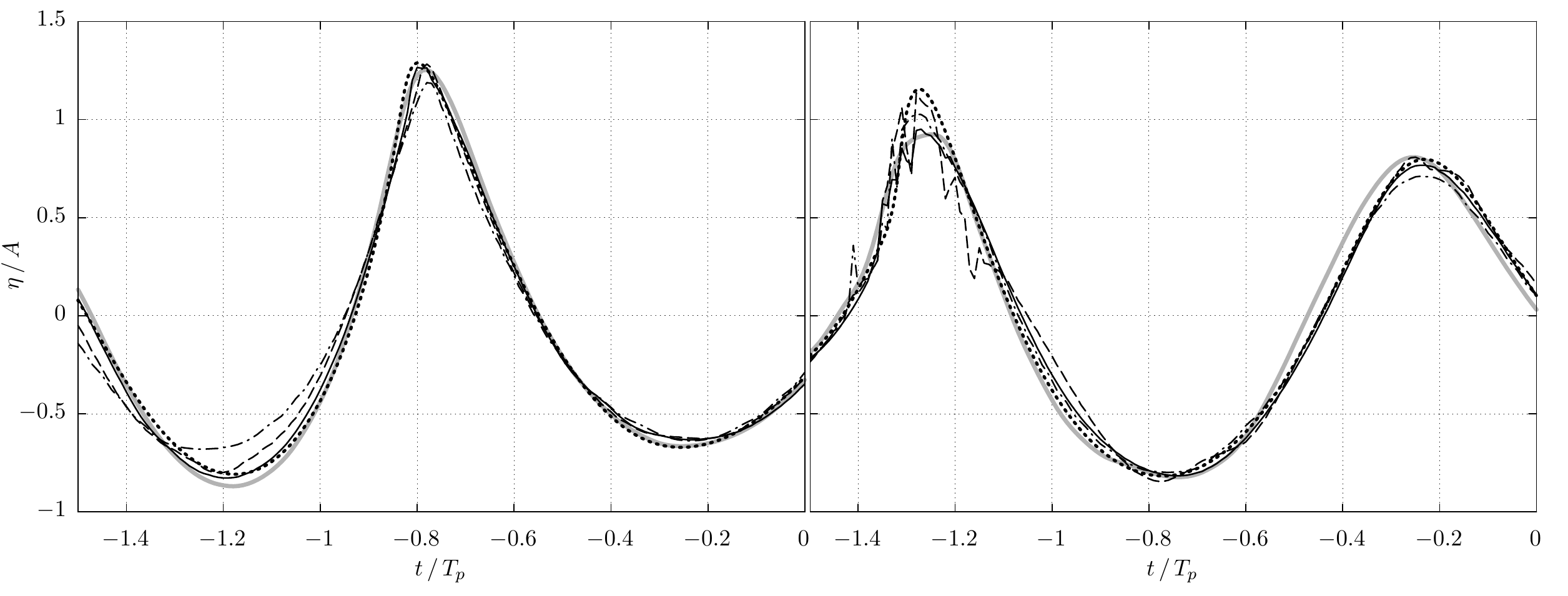}}
\caption{%
Surface elevation of a wave group at $x=0$. $A=7.5\,$cm.  Left --
	pre-breaking ($\Delta\phi=\pi$).  Right -- post-breaking
	($\Delta\phi=0$).  Top -- wave groups scale, Bottom -- details around
	the main crest / trough.  Experiment (solid grey), Lagrangian model
	(dotted), olaFlow with different mesh resolutions. Number of cells:
	$N=26832$ (dash-dotted); $N=105625$ (dashed); $N=413750$ (solid).
} 
\alabel{fig:cfd}
\end{figure}
%
%

	\subsection{Model validation and convergence}

The experiments are also reproduced in two dimensions using the VoF numerical
wave tank olaFlow. The numerical setup is identical to that described in
Section~\ref{sec:LagrangianValidation} for the Lagrangian model. The wave flume
is $12.5$\,m long and $0.56$\,m high. Waves are generated on the right boundary
with a piston-type wavemaker \cite{Higuera2015}. The wavemaker excursion has
also been reduced by $18.5\%$ to account for losses occurring on the gaps
between the paddle and the lateral and bottom walls of the experimental flume.
The bottom wall of the flume carries a no-slip boundary condition (BC). Waves
are absorbed actively at the boundary opposite to the wavemaker using boundary
conditions described in \cite{Higuera2013}. Since CFD models excel in
simulating nonlinear and wave breaking conditions, only the highest wave
conditions ($A=7.5$\,cm) were selected for two phase shifts $\Delta\phi=0$ and
$\Delta\phi=\pi$.

Three meshes with different cell size have been tested in the convergence
study. The most detailed mesh has a base cell size throughout the wave flume of
$1\,$cm in the horizontal and vertical directions. A refinement region in which
cells have high vertical resolution ($2.5$\,mm) has been defined between $z = -0.08$\,m
and $z = 0.10$\,m (see Figure~\ref{Layout}) to enhance the level of detail
along the free surface. The resulting mesh is structured and non-conformal, and
has a total of $N=413750$ square cells. The two other meshes have been built in
a similar way, but with a maximum resolution half ($5$\,mm) and one fourth
($1$\,cm) of the cell size of the previous mesh. These configurations yield
$N=105625$ and $N=26832$ cells, respectively. In the CFD simulations the time
step is not fixed, but dynamically calculated to fulfil a maximum Courant
number ($C_{o}$) of 0.15. All simulations have been run with the $k$-$\omega$
SST (shear stress transport) turbulence model defined in \cite{devolder18},
which is especially suitable for wave simulations. The computational cost of
each simulation varies due to the number of cells, which conditions the $C_o$,
and physics involved (e.g. wave breaking). The simulations of 20~seconds in
real time for the $N=26832$ mesh takes between 2.8 and 4.9 hours in a single
core. The simulations for the $N=105625$ mesh take between 23.1 and 25.4 hours
in two cores. The simulations for the $N=413750$ mesh take between 66.3 and
66.9 hours in four cores.

%
%
\begin{figure}[t!]
\centering
\includegraphics[width=0.9\textwidth]{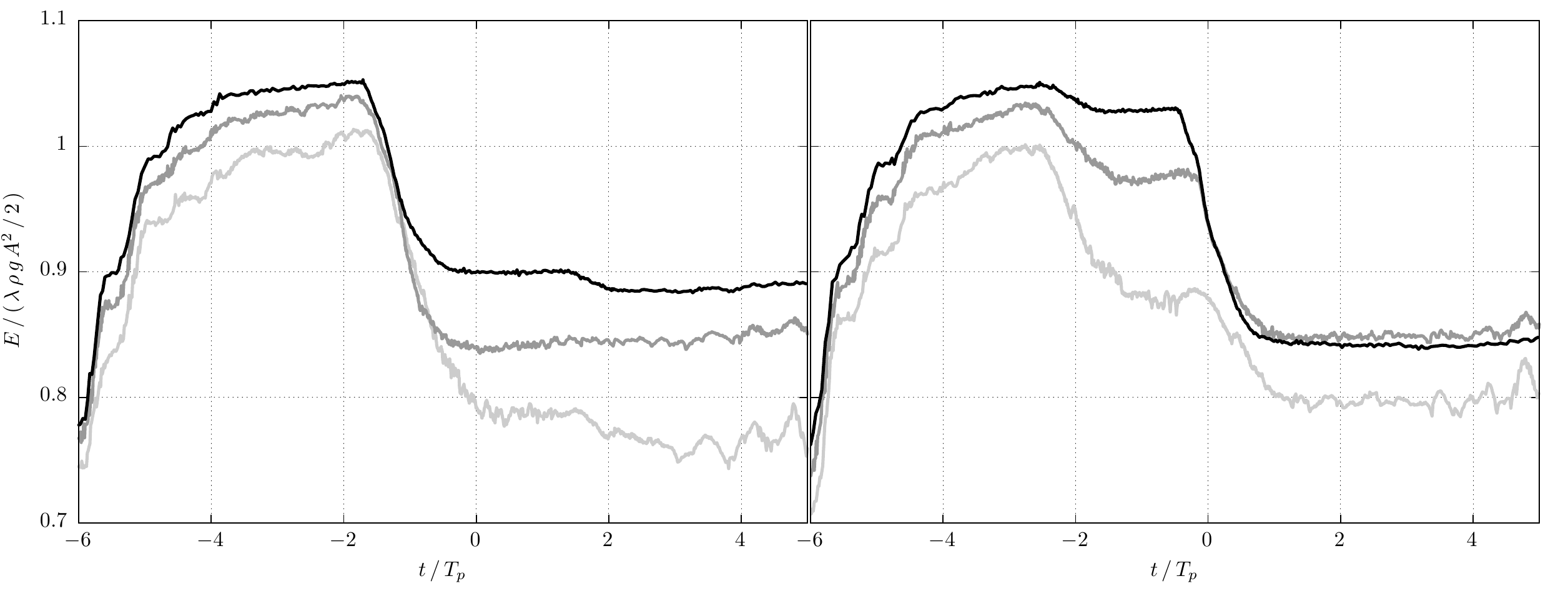}
\caption{%
Normalised total energy of breaking waves calculated by olaFlow with different
	number of cells.  $A=7.5$\,cm.  Left -- phase shift $\Delta\phi=0$.
	Right -- phase shift $\Delta\phi=\pi$.  $N=26832$ (light grey);
	$N=105625$ (grey); $N=413750$ (black).
} 
\alabel{fig:engconvVOF} 
\end{figure}
%
%

The convergence for olaFlow is evaluated in terms of the free surface
elevation (Figure~\ref{fig:cfd}) and the normalised total energy of the wave
flume (Figure~\ref{fig:engconvVOF}).  Figure~\ref{fig:cfd} presents the free
surface elevation at the centre of the tank $x=0$.  Overall it can be seen that each
refinement level in the CFD mesh yields a significant improvement in terms of
agreement with the experimental results.  The dash-dotted line corresponds to
the lowest resolution ($N=26832$), and yields the maximum deviations with
respect to the experiments.  Furthermore, the coarsest mesh produces wave
breaking artificially (top left panel, $t/T_p = -1.75$), as it is not present
neither in the experiment nor in the refined cases.  Additionally, wave
breaking occurs significantly ahead of time for the $\Delta\phi=\pi$ case
(bottom left panel, $t/T_p = -1.4$) for the lowest resolution.  The mid and
high resolution cases do not present such problems.  In fact, the solid line
solution captures remarkably well the complete profiles of the wave groups.

The total wave energy in Figure~\ref{fig:engconvVOF} has been obtained from a
slightly different numerical setup than described above.  Instead of absorbing
the waves at the left end of the wave flume, the boundary has been replaced by
a wall (no-slip BC) to prevent energy flowing out of the system, which would
affect the energy calculations.  Nevertheless, the reflected waves do not
produce significant effects due to the length of the flume and short duration
of the experiments.  The total energy has been calculated as the sum of
potential and kinetic energy, normalised by the energy of one wavelength of a
linear regular wave, with the frequency and the amplitude equal to the peak
spectral frequency and the linear focus amplitude of the wave group.  Both
potential and kinetic energy have been calculated as a cell-wise numerical
integration over those cells within the domain that contain water ($\alpha >
0.05$).

All the simulations in Figure~\ref{fig:engconvVOF} show oscillations on the
initial ramp.  These are created by the moving wavemaker, which is introducing
energy to the system.  Another common feature is that the cells resolution
plays a significant role in reaching higher levels of energy before wave
breaking.  In the left panel ($\Delta\phi=0$), the three meshes predict wave
breaking consistently after reaching the maximum value in total energy.  In
view of the maximum energy level and the energy level of the plateau after wave
breaking, it can be observed that the higher the resolution, the lower the
energy dissipation due to breaking.  In the cases shown in the right panel
($\Delta\phi=\pi$), this last conclusion only holds initially, for the wave
breaking event at $t/T_p = -2$.  However, the main breaking event at $t/T_p =
0$ shows the opposite behaviour, with increasing dissipation with increasing
resolution.  Generally, the dissipation in the leading breaking event decreases
with higher resolutions, but it can increase in the following event because the
wave loses less energy in the first event and remains more energetic when the 
second event takes place.

Overall, the results from the convergence tests presented in this section show
a significant improvement for smaller cell sizes when compared against the
experimental results. Also, the observations made are aligned with those in
\cite{vyzikas2018}. In their work they demonstrated that insufficient
resolution and large $C_o$ lead to over-predictions of wave amplitude, leading
to premature and more pronounced wave breaking \cite{vyzikas2018}.

%
%
\begin{figure}[t!]
\centering
\includegraphics[width=0.9\textwidth]{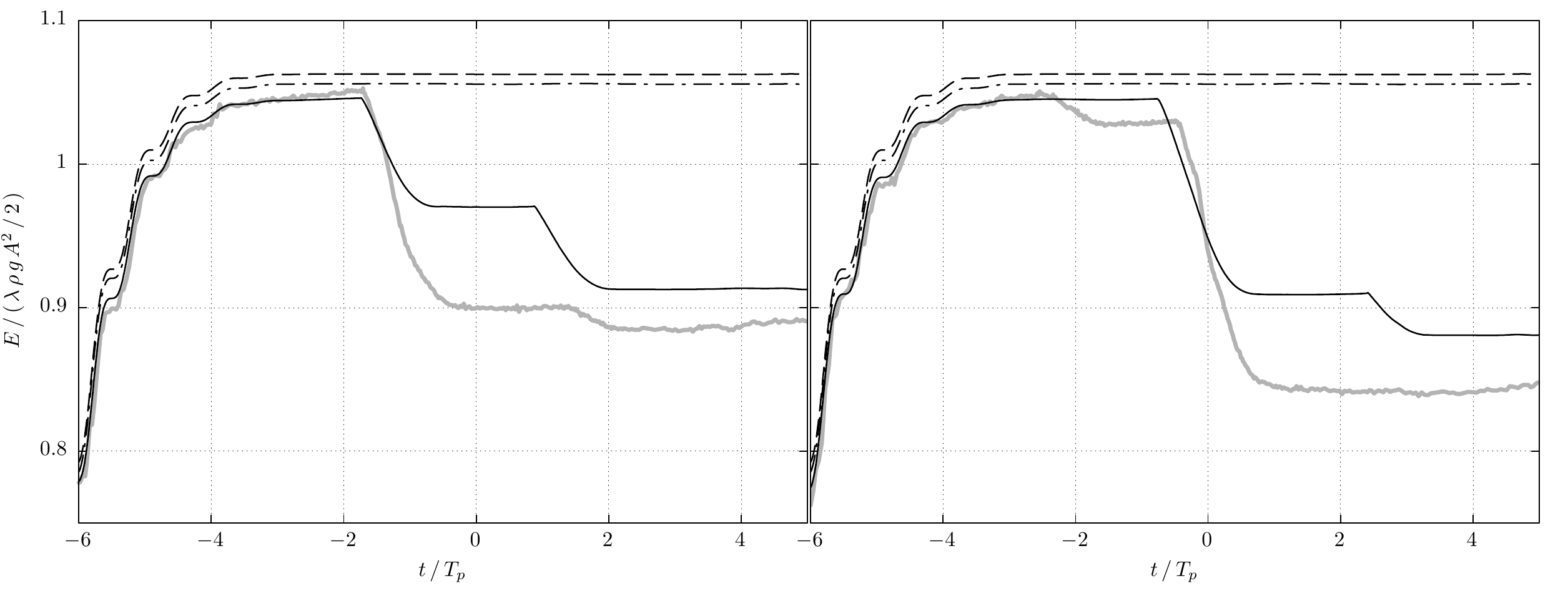}~~
\caption{%
	Normalised total wave energy for Lagrangian (black) and VoF (grey) solvers.
	Left -- phase shift $\Delta\phi=0$.
	Right -- phase shift $\Delta\phi=\pi$.
	$A=2.5$\,cm (dashed); $A=5$\,cm (dash-dotted) and $A=7.5$\,cm (solid).
}
\alabel{fig:eng}
\end{figure}
%
%

\section{Results}%

The olaFlow model and the Lagrangian model are applied in this section to
simulate the evolution of a steep wave group with a focus on modelling the wave
breaking process. The calculations by the Lagrangian model are performed using
a $401\times 16$ mesh and the time step $\delta t=0.0025$\,sec. The mesh with
413750 cells and $C_o = 0.15$ is used for the \of{} computations.
Figure~\ref{fig:eng} shows the evolution of the total wave energy for amplitude
$A=7.5$\,cm and phase shifts $\Delta\phi=0$ and $\Delta\phi=\pi$, which
correspond to inverse wavemaker input signals. Plots for amplitudes of
$A=2.5$\,cm and $A=5$\,cm calculated with the Lagrangian model are presented as
a reference. The energy is normalised as described in
Section~\ref{sec:LagrangianValidation}. The wave energy increases from the
initial zero level after the wavemaker starts operating. After the wavemaker
stops at $t/T_p\approx-3$, the total energy remains constant until wave
breaking occurs. As expected, the energy of non-breaking waves is the same for
both phase shifts.

For $\Delta\phi=0$ the Lagrangian model predicts two breaking events of similar
intensity, which are symmetrical with respect to $t=0$ and, therefore, with
respect to the centre of the tank. This qualitative behaviour is confirmed by
experimental observations. In contrast, olaFlow predicts a strong initial
breaking event and a small secondary breaking event. For $\Delta\phi=\pi$, the
Lagrangian results demonstrate an intensive breaking event near the centre of
the tank close to $t=0$ and a smaller breaking event near $t/T_p=2.5$.
The olaFlow model presents a small breaking event at
$t/T_p\approx-2.5$ prior to the main one. Only one major breaking
event in the centre of the reservoir was observed in the experiment in this
case. For both cases the VoF model predicts higher overall energy dissipation.
The convergence results of the VoF model for $\Delta\phi=0$ presented in
Figure~\ref{fig:engconvVOF} show that the increase in mesh resolution results
in decreasing intensity of the main breaking event and the development
of a smaller second breaking event. For $\Delta\phi=\pi$, the higher resolution
results in the decreasing intensity of the first breaking event and the
development of a strong breaking event at $t = 0$. It should be noted, that VoF
breaking results are not fully converged. It is reasonable to assume that a
further increase in mesh resolution will lead to the observed qualitative
breaking behaviour with two symmetric breaking events for $\Delta\phi=0$ and a
single strong break event at $t = 0$ for $\Delta\phi=\pi$. Given the
computational time required to obtain these results, it was found impractical
pursuing an additional level of resolution.  For the Lagrangian solver, the
difference with the observed breaking behaviour is due to the application of
the nonphysical heuristic breaking model, which requires an adjustment of the
parameters to obtain the best performance for each individual case. Detailed
calibration of the Lagrangian model for different breaking conditions is
outside the scope of the present paper.

In the reminder of this section we perform a detailed comparison of the
behaviour of the two models around the breaking crest. We focus our attention
to the large breaking event at the centre of the tank and consider the wave
with $A=7.5\,$cm and $\Delta\phi=\pi$. Figure~\ref{fig:crst} shows the time
history of wave crest elevation at three locations along the tank. The
corresponding wave profiles can be seen in Figure~\ref{fig:prf1}. The
difference between measured and calculated wave crests can be observed in
Figure~\ref{fig:crst} near the top of the crest. It should be noted that for
very steep wave peaks the experimental measurements by wave probes are
unreliable. This is due to the high-speed flow at the crest, which creates a
cavity around the wires of the wave probe, resulting in a reduction in the
recorded crest elevation. However, for the main part of the crest, the
calculated and experimental results are compared with good accuracy. The
rounded end of the overturning wave observed in Figure~\ref{fig:prf1} is
explained by surface tension, which produces a considerable effect given the
relatively small experimental scale. For the initial phase of wave breaking,
there is a very good agreement between the Lagrangian model without the
breaking model and olaFlow. However, olaFlow predicts a slightly faster
development of the breaker. These differences can be caused by many reasons
such as insufficient mesh resolution at the crest for the VoF model or the
exclusion of air effects by the Lagrangian model. Since the breaking process is
extremely sensitive to small disturbances, the achieved level of comparison can
be considered satisfactory.

%
%
\begin{figure}[t!] \centering
\resizebox{0.9\textwidth}{!}{\includegraphics{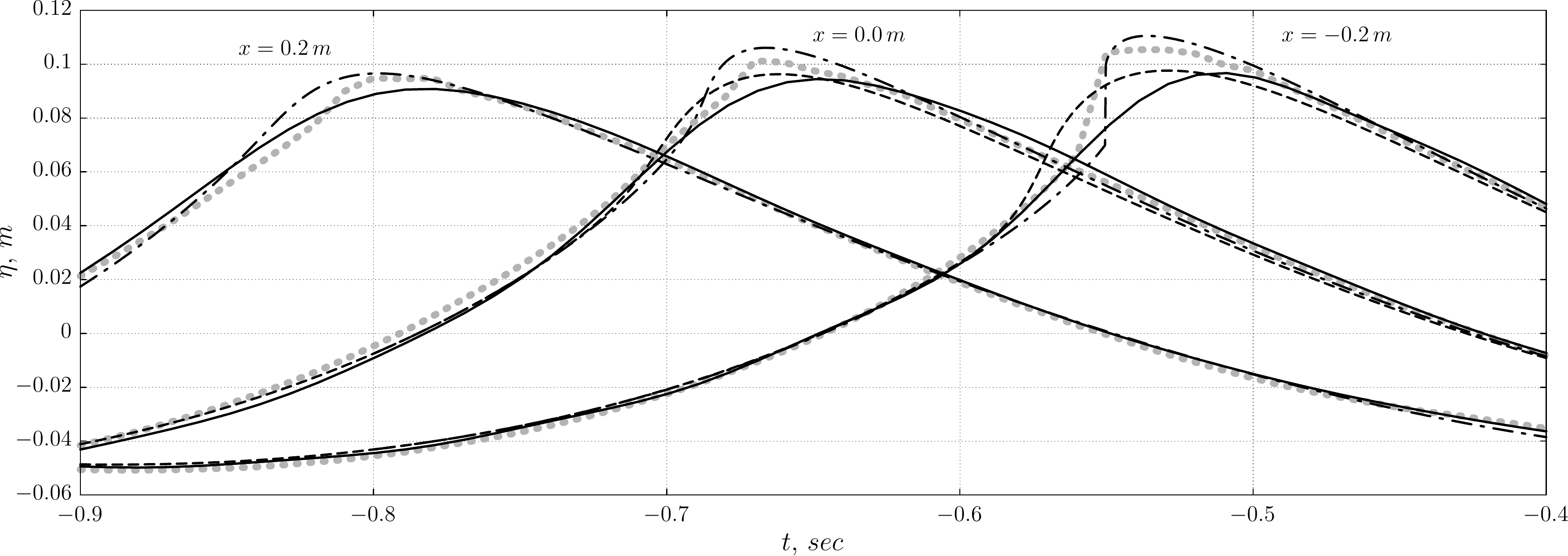}~~~}
\caption{%
Time histories of wave crest evolution at different positions along the flume,
	$A=7.5$\,cm, $\Delta\phi=\pi$.  Experiment (solid); VoF model (grey
	points); Lagrangian model (dash-dotted); Lagrangian model with breaking
	model (dashed).
} 
\alabel{fig:crst} 
\end{figure}
%
%

%
%
\begin{figure}[t!]
\centering
\resizebox{\textwidth}{!}{\includegraphics{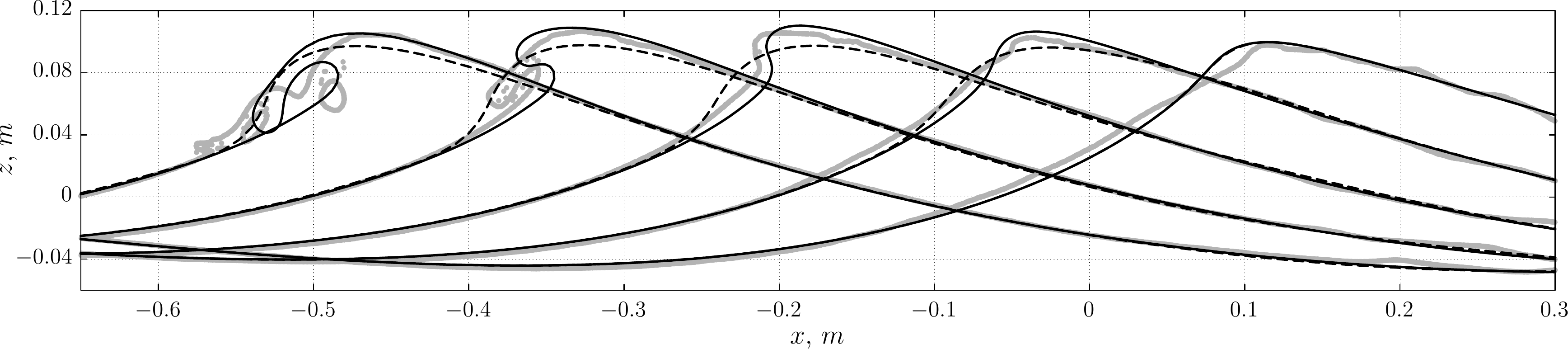}}
\caption{%
	Breaking crest evolution between $t=-0.76$\,sec and $t=-0.36$\,sec.
	VoF model (thick grey);
	Lagrangian model (solid);
	Lagrangian model with breaking model (dashed).
}
\alabel{fig:prf1}
\end{figure}
%
%

For the wave peak at $x=0.2\,$m, the Lagrangian breaking model is not yet
activated and thus the Lagrangian solutions with and without the breaking model
are identical. At $x=0$, the top of the wave crest begins to deform. It moves
faster than the main body of the crest and eventually overturns. At this stage,
the damping term of the breaking model is activated, as illustrated by the
differences in the crest shape with and without the breaking model. Further in
the tank, at $x=-0.2\,$m, the top of the crest begins to overturn and a
vertical front develops. The shape of the crest computed with the active
breaking model is seen to differ considerably from the shape of the
freely-evolving crest. However, the main differences are located near the top
of the crest, while other parts of the wave are unaffected by the breaking
model. Calculations without the breaking model are continued until the self
contact of the overturning wave at $t=-0.36\,$sec (Figure~\ref{fig:prf1}).
After that, the results produced by the Lagrangian model without the breaking
model are not physically meaningful. The breaking model, on the other hand,
does not show overturning, but allows further calculations beyond this point.

In the last two profiles ($t=-0.46\,$sec and $t=-0.36\,$sec), olaFlow shows
forming of a large air pocket and smaller bubbles. The model also predicts the
wave crest bouncing from the free surface of the main water body, creating a
second plunger. However, it must be noted that the olaFlow simulation is 2D,
whereas wave breaking is typically a highly 3D process. This means that, for
example, trapped air will only be able to escape upward, whereas in 3D it might
also escape sideways. In addition, the mesh resolution achieved in olaFlow
calculations does not provide full convergence of breaking results. Therefore,
the present solution must be regarded with the limitations inherent to 2D
modelling and insufficient mesh resolution. Unlike the VoF model, the
Lagrangian model is not able to simulate the air entrapment. However, as can be
seen in Figure~\ref{fig:prf1}, the solution with the breaking model reproduces
the shape of the breaking wave crest relatively well, and provides a good
starting point for subsequent simulations of the post-breaking wave.

Examples of horizontal and vertical velocity profiles at the wave crest and on
the front and rear slopes are given in Figure~\ref{UW}. The profiles are
presented at the moment when surface elevation reaches its maximum at $x=0$ and
the crest starts overturning. There is a good general agreement
between the profiles calculated by olaFlow and the Lagrangian model without the
breaking model. The higher peak velocity for the VoF model is consistent with
the previous observation of faster crest deformation. The difference between
the velocities calculated by the Lagrangian model with and without the breaking
model can be observed near the top of the wave crest but disappears rapidly
everywhere else. It is important to note the good degree of agreement between
velocity profiles and surface elevation calculated by the Lagrangian model with
the breaking model and olaFlow for the region outside the immediate vicinity of
the breaking crest.

%
%
\begin{figure}[t!] \centering
\resizebox{0.9\textwidth}{!}{ 
	\includegraphics{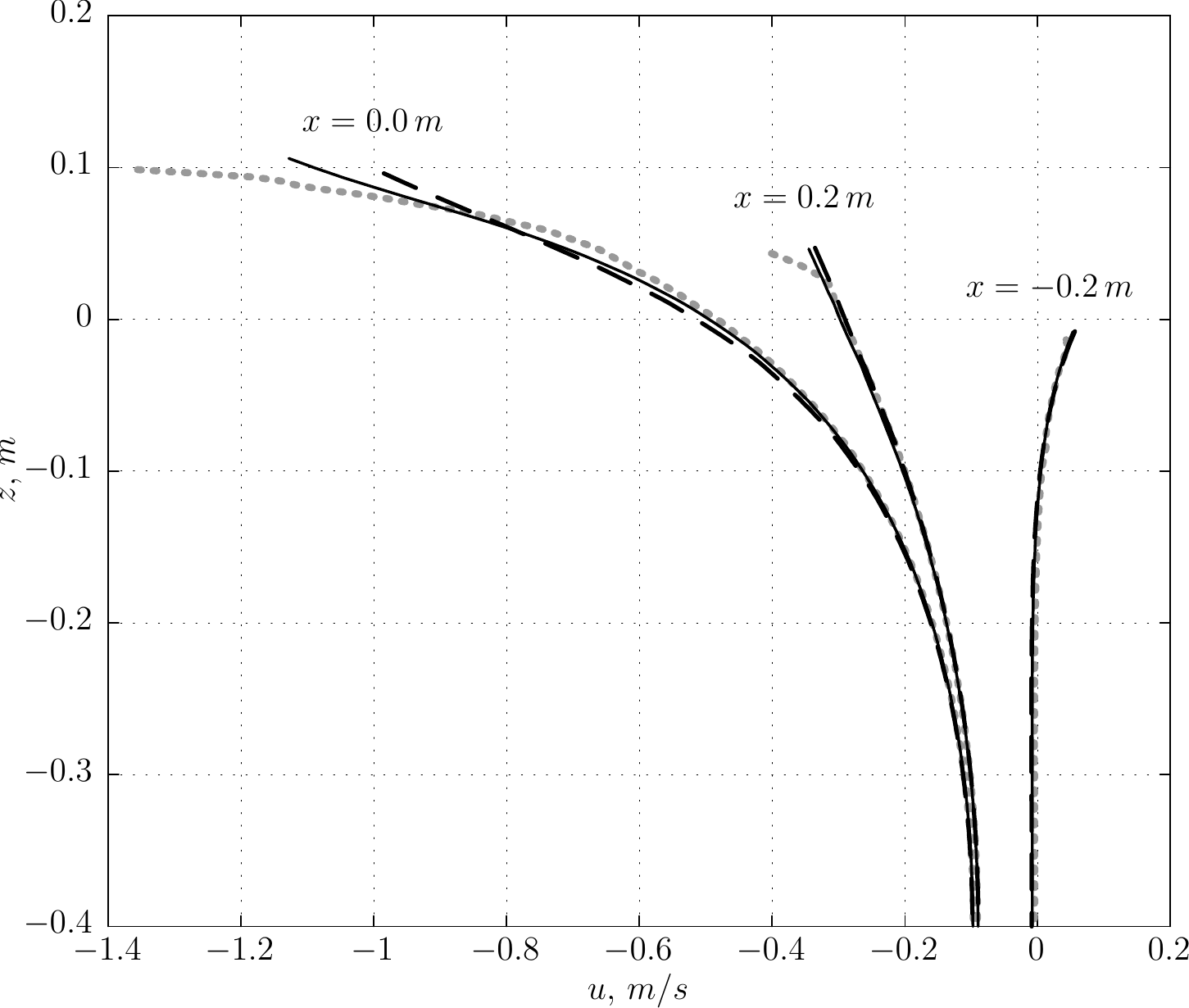}
	\includegraphics{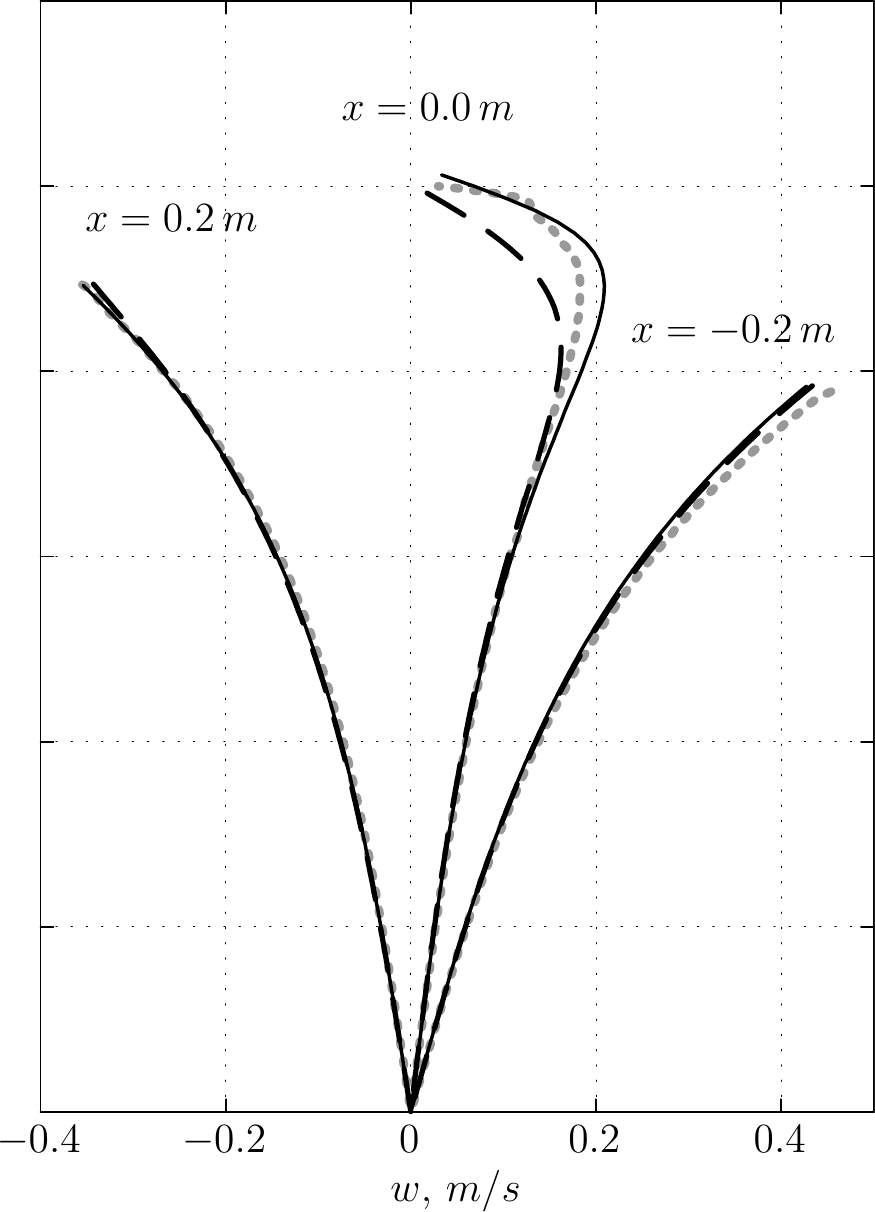}~~~~~~~~~~ }
\caption{%
	Horizontal (left) and vertical (right) velocity component profiles at
	$t=-0.665$ and different positions along the flume. VoF model (dotted
	grey); Lagrangian model (solid); Lagrangian model with breaking model
	(dashed). The time corresponds to the maximal surface elevation at
	$x=0$.
} 
\alabel{UW} 
\end{figure}
%
%

\section{Discussion and conclusions}%

The main result of the paper is the development of a fully Lagrangian numerical
model and its application to the evolution of steep wave groups with breaking.
The model is able to simulate the long-term evolution of highly nonlinear waves
and to calculate wave profiles and kinematics close to spilling breaking. A
4-parametric dissipative breaking model is introduced to prevent the breakdown
of computations, which usually follows the breaking onset. To improve the
accuracy of modelling of the long-term evolution of travelling waves, a
dispersion correction term is introduced to the free surface boundary
condition, which increases the order of approximation of the dispersion
relation. 

The results of Lagrangian calculations for surface elevation are validated
against wave flume experiments and for surface elevation and kinematics against
numerical results obtained by a VoF model. OlaFlow \of{} realisation of VoF
with a recent addition of a realistic method of wave generation with a piston
type wavemaker \cite{Higuera2013,Higuera2015} is used in this paper.
Although this may seem trivial, before introducing this addition, comparing
\of{} computations for water waves with experiments and results of other
numerical models was not an obvious task, mainly because of the different
methods of wave generation. After introducing a realistic wavemaker, it became
possible to create an \of{} wave tank replicating the actual experimental wave
flume. Since the same experimental flume is simulated by the \of{} model and
the Lagrangian solver, the direct comparison of the experimental and numerical
results becomes straightforward. 

For a steep non-breaking wave group, both the \of{} results and the results of
the Lagrangian model with dispersion correction demonstrate a perfect match
with experimental surface elevation measurements. An equally good comparison
up to breaking onset is also reported for the steepest wave group tested. Post
breaking behaviour of the wave group is not simulated that well by either model
(Figure~\ref{fig:cfd}). For the Lagrangian solver this is caused by the
application of the nonphysical heuristic breaking model, which requires
selection of model parameters. For the VoF solver, this is mainly due to an
insufficient mesh resolution in the vicinity of the breaking crest. Another
reason is the application of a two-dimensional mesh, which does not allow to
resolve the three-dimensional details of the formation and the evolution of the
turbulent breaking roller. It should be noted that for both models, the
difference is mainly concentrated around the breaking crest and that the
overall evolution after the breaking is simulated relatively well.

Further comparison between the numerical results is performed for parameters
not measured in the experiment. Simulations of an overturning crest with
olaFlow and the Lagrangian model without the breaking model demonstrate a good
agreement (Figures~\ref{fig:crst},~\ref{fig:prf1}). For the Lagrangian model
with the breaking model the difference of the profiles is localised in a small
area around the crest and disappears for the rest of the wave profile. The same
can be said about velocity profiles under the crest just before the breaking
(Figure~\ref{UW}). There is a good agreement between velocity profiles for the
VoF model and for the Lagrangian model without the breaking model. For the
Lagrangian model with suppressed breaking, the difference between the
calculated velocities can only be observed in the vicinity of the crest.
Overall, when compared with olaFlow computations of wave breaking, the results
of Lagrangian calculations with the breaking model demonstrate plausible
behaviour with intensive energy dissipation (Figure~\ref{fig:eng}).

%
%
\begin{figure}[t!]
\centering
\resizebox{0.85\textwidth}{!}{\includegraphics{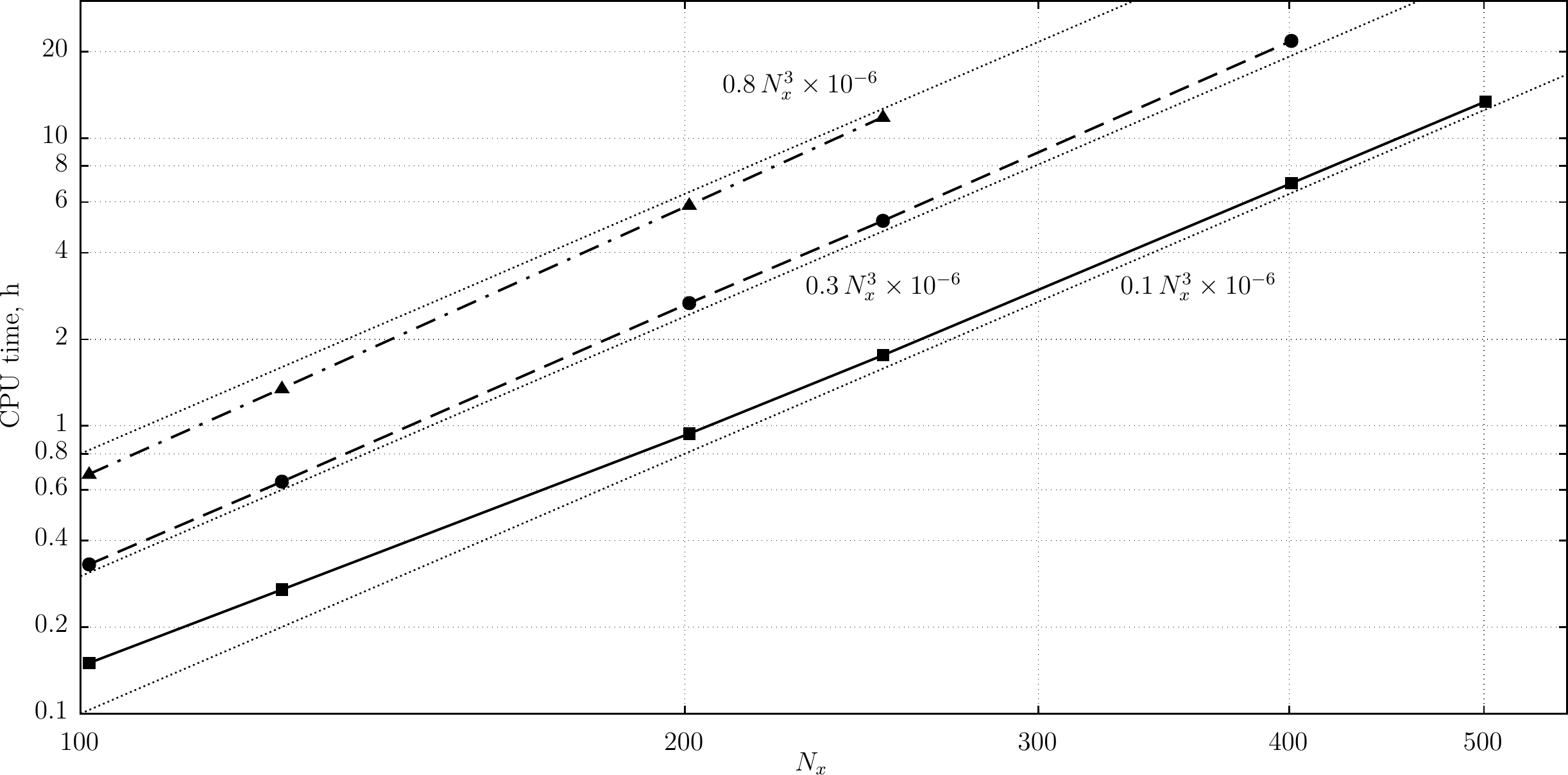}~~~~}
\caption{%
Computation time for modelling of 20\,sec of wave propagation 
by the Lagrangian solver on a single 2.4\,GHz CPU for different mesh size:
$N_z=11$ (solid); $N_z=16$ (dashed); $N_z=21$ (dash-dotted).
}
\alabel{fig:CPU}
\end{figure}
%
%

Computational efficiency of the Lagrangian solver is important for
its applicability to large-scale problems, especially in 3D. In the implicit
time marching scheme implemented in the Lagrangian solver, the most time
consuming element is the inversion of a Jacobi matrix used by the Newton
iterations for solving nonlinear grid equations. The required calculation time
grows fast with increasing matrix dimension, which is proportional to the
square of the product of dimensions of a numerical mesh. For a matrix
inversion algorithm used in this work, the inversion time is approximately
proportional to the square of the dimension of the matrix. For a 2D problem
with a constant value of $N_z$ much smaller than $N_x$, the matrix dimension is
proportional to $N_x$. To provide a uniform discretisation error in space and
time, the value of time step should be proportional to $\delta x$, which implies
that the number of steps is proportional to $N_x$. Thus, the overall
calculation time grows with a rate proportional to $N_x^3$, as illustrated by
Figure~\ref{fig:CPU}. For an explicit method, like the one used in olaFlow, the
computational time estimated in the same way is proportional to $N_x^2$. Due to
better numerical stability, the Lagrangian model uses considerably larger time
step. Moreover, it requires less horizontal points to resolve the free surface
since the latter is well defined by the upper boundary of the Lagrangian
computational domain. As a result, the Lagrangian model requires much less
spatial points and less time steps to obtain a stable solution of the same
accuracy as the \of{} model. 

If the number of mesh points increases due to a larger computational domain or
high mesh resolution, the computational time for the Lagrangian model grows
much faster than that for the \of{} model. For a sufficiently large number of
points, the Lagrangian model loses its advantage in computational efficiency.
This disadvantage of the Lagrangian model becomes crucial for 3D problems,
which restricts the applicability of the current version of the Lagrangian
solver. High demand for computational resources for large scale problems is a
well recognised disadvantage of implicit schemes, which often outweighs their
advantages in numerical stability. The radical method of increasing
computational efficiency is using parallel computing. Implicit solvers, like
the one used in the Lagrangian model, have a single standard time-consuming
operation and are therefore suitable for efficient parallelisation. A parallel
version of an implicit solver can be created with minimal changes to the
original code by replacing a matrix inversion subroutine with a parallel
analogue. This feature is particularly useful in light of recent advances in
GPU-based matrix inversion algorithms, which are much faster than conventional
parallelisation using multiple CPUs \cite{Sharma2013}.

The good agreement between the velocity profiles and the surface elevation
calculated by the Lagrangian model and the olaFlow model makes it possible to
use them to create a hybrid model in order to reach a new qualitative level of
efficiency and accuracy of modelling. The elements of a hybrid model are used
to simulate different parts of the overall process. The Lagrangian model more
efficiently simulates long-term wave evolution than the VoF model. However, a
non-physical breaking model is needed to continue the calculations after wave
breaking. On the other hand, the VoF model is able to directly simulate the
breaking process, but a high mesh resolution is required. It is therefore
advantageous to create a hybrid model combining the Lagrangian model to
simulate the wave group evolution and the VoF model to simulate the details of
the breaking process. More specifically, the VoF model to be used to model the
small region around the breaking crest, which allows to obtain a high mesh
resolution with a relatively small number of mesh points and therefore with a
low computational cost. As a first step, the Lagrangian and VoF coupled models
were recently applied to fluid-structure interaction problems, which requires
simpler coupling techniques, compared to a hybrid breaking model
\cite{Higuera18,Chen2019}.

In summary, the results presented in the paper demonstrate high quality
validation for both numerical models. The Lagrangian is able to simulate the
evolution of steep wave groups before and after breaking. The accuracy achieved
by the Lagrangian model is comparable to that achieved by a VoF model but the
required computational resources are considerably smaller. The level of
agreement between the surface elevation and the velocities around a breaking
crest, calculated by both models, makes them promising elements of a more
efficient hybrid model.


\end{document}